\documentclass{sig-alternate}
\pdfoutput=1

\newcommand{\mypar}[1]{{\bf #1.}}
\newtheorem{obs}{Observation}

\usepackage[colorlinks=true,linkcolor=blue,citecolor=blue]{hyperref}

\newcommand{\doublefigure}[5]{{\begin{figure}[hbt] %
\centering %
\includegraphics[width=#1\linewidth]{#2}\\
\includegraphics[width=#1\linewidth]{#3} %
\caption{ #4}%
\label{#5}%
\vspace*{-6pt}
\end{figure}}}

\newcommand{\Doublefigure}[5]{{\begin{figure*}[t] %
\centering %
\includegraphics[width=#1\linewidth]{#2}
\includegraphics[width=#1\linewidth]{#3} %
\caption{#4}%
\label{#5}%
\vspace*{-6pt}
\end{figure*}}}
 
\newcommand{\singlefigure}[4]{{\begin{figure}[htb] %
\centering %
\includegraphics[width=#1\linewidth]{#2}\\
\caption{#3}%
\label{#4}%
\vspace*{-6pt}
\end{figure}}}

\newcommand{\A}{\sigma}
\newcommand{\SA}{U_{\sigma}}
\newcommand{\FA}{F_{\sigma}}
\newcommand{\Ai}[1]{{\bf \sigma}_{#1}}
\newcommand{\B}{\pi}
\newcommand{\SB}{U_{\pi}}
\newcommand{\FB}{F_{\pi}}
\newcommand{\Bi}[1]{{\bf \pi}_{#1}}

\newcommand{\Jaccard}[2]{{\bf J}(#1,#2)}
\newcommand{\DELTA}[2]{{\bf \delta_{term}}(#1,#2)}

\graphicspath{{./}}

\begin{document}
\hypersetup{pdfauthor={Paolo D'Alberto},pdftitle={Search Engine Correlation}}


\CopyrightYear{2010}

\crdata{} 


\numberofauthors{2}


\title{On the Weakenesses of Correlation Measures used for Search
  Engines' Results} \subtitle{(Unsupervised Comparison of Search
  Engine Rankings)}


\author{
\alignauthor 
  Paolo D'Alberto \\ 
  \affaddr{Yahoo! Inc.}\\ 
  \affaddr{Sunnyvale, CA, USA}\\
  \email{pdalbert@yahoo-inc.com}
\alignauthor 
  Ali Dasdan\\ 
  \affaddr{Knowledge Discovery Consulting}\\ 
  \affaddr{San Jose, CA, USA}\\
  \email{ali\_dasdan@yahoo.com}
} 

\maketitle
\begin{abstract}
  The correlation of the result lists provided by search
  engines is fundamental and it has deep and multidisciplinary
  ramifications. Here, we present automatic and unsupervised methods
  to assess whether or not search engines provide results that are
  comparable or correlated. We have two main contributions: First, we
  provide evidence that for more than 80\% of the input queries
  ---independently of their frequency--- the two major search engines
  share only three or fewer URLs in their search results, leading to
  an increasing divergence. In this scenario (divergence), we show
  that even the most robust measures based on comparing lists is
  useless to apply; that is, the small contribution by too few common
  items will infer no confidence.  Second, to overcome this problem,
  we propose the fist content-based measures ---i.e., direct
  comparison of the contents from search results; these measures are
  based on the Jaccard ratio and distribution similarity measures (CDF
  measures). We show that they are orthogonal to each other (i.e.,
  Jaccard and distribution) and extend the discriminative power
  w.r.t. list based measures.  Our approach stems from the real need
  of comparing search-engine results, it is automatic from the query
  selection to the final evaluation and it apply to any geographical
  markets, thus designed to scale and to use as first filtering of
  query selection (necessary) for supervised methods.
\end{abstract}

\section{Introduction}
\label{sec:intro} 

Today users have access to many search engines providing services for
their web search needs but the top three search engines attract almost
all user queries and the top search-engines provide service to more
than two-thirds of the search traffic (as today 95\%). What is the
reason for this situation? Attempting to answer this question and
other similar questions, prompted us to the study of the metrics for
comparing search engines. Many such metrics are already available,
such as relevance, coverage, and presentation (e.g., see the tutorial
\cite{DasdanTV2009}).  Independent of the metric, we would expect
that, given the same query, if two different search engines return
results that are similar in both contents and order, then the users'
satisfaction should be similar.  In this work, we argue that the
previous hypothesis (i.e., similar results) can be measured; the
conclusion (i.e., user satisfaction) is more subjective and we show
that we must have a supervised approach. 

We also show that a leading search-engine is not always (and should
not be always considered as) the ultimate reference of users'
satisfaction nor quality. \footnote{If a leading search engine, given
  a query, provides a set of URLs, we do not suggest to provide the
  same set to any engines. }

Thus, how can we judge the similarity of two sets of search results?
By representing URLs as sets or lists, we do take advantage of these
measures: For example, we can use the Jaccard ratio for set similarity
(without confidence level), we can use Spearman's footrule and
Kendall's tau for list similarity (with confidence level and for lists
that are permutations and without weights). However, different search
engines provide results that are never permutations, at best, are
sparse lists, and the URLs should not be treated equally because users
pay attention only to the top results (pay little attention to the
bottom results, skip the successive result pages and just refine the
query). These measures, in combinations with adaptations for sparse
lists, are still the state-of-the-art measures and they are the first
we used.

As we show in this work, for more than 80\% of the queries the overlap
between two sets of search results is less than 30\%.  Unfortunately,
This observation implies that the top search engine does not subsume
the results returned by the next major search engine and URL-based
measures are insufficient for comparing different search engines with
such a little overlap. But why this small overlap affect the quality
of URL-based measures? Intuitively and in practice, these measures
work well on the common URLs quantifying their difference but the
no-common URLs dilute the measure making them less and less sensitive.

We show in this work that when the overlap is low between the results
of two search engines, the relative quality (users' satisfaction)
between search engines varies widely. We looked at the correlation
between the URL overlap (Jaccard) and the quality of the search
results measured by the discounted cumulative gain (DCG)~\cite{JaKe02}
(which is a supervised measure because is an editorial--human
measure). We have found that the results vary widely in quality
especially when the overlap is low: this implies that any search
engine can return better or worse results depending on the query and
it is difficult to estimate the outcome reliably.  But, once more, why
this small overlap affect the quality of URL-based measures?  Most of
the queries will provide uncorrelated values: we must use instead
precious human resources to distinguish the queries that provide
different results (i.e., if we could measure the queries that provide
similar results, we may infer similar users' satisfaction).

We show in this work that content-based similarity measures provide
more discriminating conclusions than URL-based similarity measures.  A
URL is nothing more than a pointer where the information is. The
contents must be interpreted and quantified as we summarize in the
following paragraph:

We propose to use the contents from search results landing pages for
computing similarity. In particular, we represent the contents by a
set of terms as well as a distribution of terms and adapt the Jaccard
ratio and many distribution-similarity measures from
\cite{DAlbertoD09} (we present results for the extension of the $\phi$
measure \cite{KiferBG2004} in particular to compute similarity of
free-format documents).  Ultimately, contents based measures
outperform lists based measures when applied in an unsupervised
fashion.

As practitioners of pairwise correlation measures for search engine
comparison and similarity computation, we are aware that
rank correlation of search engines is used as common example or
flagship for the application of list-based correlation measures.  We
want to make aware the community that there are more sophisticated
measures.

The rest of the paper is organized as follows. We introduce the
related work in \S~\ref{sec:related} and a theory of similarity in
\S~\ref{sec:theory}. In \S~\ref{sec:practice}, we present how the
theory is applied in practice to our choice of similarity measures and
their parameters. We present the experimental methodology in
\S~\ref{sec:method} and the experimental results and our observations
in \S~\ref{sec:experimental-results}. We conclude in
\S~\ref{sec:conclusions}.

\section{Related Work}
\label{sec:related}

In the following, we will attempt to present a
representative though limited set of related works in the fields of
list correlation, coverage and similarity measures (the three
components of our method). As such, we introduce previous results in
the context of our work in such a way to present the main differences
and then useful references for a deeper investigation.

Correlation measures have a long history and by
nature are interdisciplinary. We can start with the contributions by
Gauss, Laplace, and Bravais; however, the first reference/introduction
to the term correlation is by Galton~\cite{Galton1888}: where it is
crystallized that the variation of two organs are due to common causes
and proposed a reversion coefficient, as also discussed by
Pearson~\cite{Pearson1920A}.

Spearman proposed the footrule in 1906~\cite{Spearman1906} with its
distribution in a psychology journal, but he turned his attention to
rank correlation (comparable rankings for addition and pitch).

Concurrently, the Jaccard ratio was introduced in
1901~\cite{JaccardOriginal1901} and used for the species-to-genus
ratio~\cite{JaccardApplication1901} as introduced in a historical note
by \cite{Jarvinen1982}. The ratio was used as measure of variety.  No
probability concept or confidence was introduced . Here, we use the
ratio in a similar spirit and without a probability distribution.

Kendall in 1938 introduced a new measure of rank
correlation~\cite{Kendall1938}, based on the count of how many swaps
of adjacent elements are necessary to reduce one list to another as in
the bubble sort algorithm. From then, different versions of
correlation measures (with and without weights) have been used and
presented (e.g., see \cite{Ta02} for a short survey). For example,
Kendal's with weights has been proposed by
Sievers~\cite{Sievers1978}. 

Rank correlation aims at the measure of disarray/concordance
especially of short permutations. Its applications range in so many
different fields and applications: medicine, psychology, wherever data
is incomplete, to capture trends, and rank aggregation (e.g., see the
reviews in \cite{DworkKNS2001,Sculley07}).

About the rank correlation and their comparison, the literature is
quite large, of the recent publications we may cite
\cite{Carterette2009} and \cite{YilmazAR2008} where the authors
introduce a new measure starting from the the Kendall's coefficient
for the information retrieval field.

Closer to our research is the comparison of search engines rankings by
Bar-Ilan et al.~\cite{Bar-IlanML2006}: The idea is to set a small set
of queries and monitor search engines ranking in time. The query set
has a relative high intersection in the result lists (common results
at least between Google and Yahoo!). In contrast, we show that our
query corpus is large and has wider variety. 

We conclude this section by citing the work by Fagin et
al.~\cite{FaKuSi03,FaKuMaSiVe04}, where they present various distance
measures for unweighted partial lists. These papers are excellent
references for partial list similarity measures, their various
generalizations, their equivalence, and some results on the comparison
of search engines. In a different work \cite{DasdanD2011}, our proof
of the equivalence for the weighted generalizations has the same
spirit as the results in these papers.

The coverage and overlapping of search engines is a new problem where
one of the first attempts to measure such a difference has been
proposed in 1998 \cite{BharatB1998}. The same paper needed a few tools
for the similarity of documents such as shingles that we still use
today. About similarity measures of documents, the literature is as
large and old as for the correlation measures and it is multifaceted:
an arbitrary classification is by signature comparison and by contents.
By signature, two documents are compared by summaries or signatures
only (e.g., see \cite{Charikar2002,Broder1997}).  We use the Jaccard
ratio of the signature because: first, it is common in the field the
authors work (e.g., see \cite{DasdanDSD09} for another use); and
second it is more a literal comparison than a semantic comparison. We
actually use a signature of up to 1000 items (shingles), thus
performing more a contents comparison than a probabilistic comparison,
reducing to zero false positives. By contents, we could use any
bag-of-words ---e.g., word--count histograms--- measures, and thus use
stochastic measures; for example, one of the first measures is
proposed by Kolmogorov in 1933, but for a recent survey see
\cite{DAlbertoD09}.


For each of these metrics, and especially for the relevance metrics,
the rank of a search result plays an important role. The reason is
that users expect to find the answer among the top search results, and
the probability of a click (i.e., the user takes a look at the page)
drops quite drastically as the rank increases. In parallel with our
work (i.e., they cited this work), Kumar and Vassilvitskii
\cite{KumarV2010}, present measures so that to take in account the
relevance of a document in conjunction with its rank. Of course,
relevance is (currently) a supervised feature.


\section{A Theory of Similarity}
\label{sec:theory}

In this section, we provide the mathematical overview of comparing
sets, lists, and distributions. Due to almost a century-old history on
the subject, our discussion is necessarily focused on the measures
that we use in this study. In the case of list similarity, we have a
contribution by providing a weighted generalization of Spearman's
footrule and Kendall's tau and prove their equivalence for
permutations and partial lists but we presented separately
\cite{DasdanD2011}. For list with little overlap, we introduce novel
metrics.

\subsection{Set Similarity}
\label{sec:set-similarity}

Given two sets $\SA$ and $\SB$, their intersection and union are
defined as
\begin{equation}
\SA\cup\SB=  \{ x | x \in \SA \text{  or } x \in \SB \}
\end{equation}
and
\begin{equation}
\SA\cap\SB = \{ x | x \in \SA \text{  and } x \in \SB \},
\end{equation}
where elements are included without repetition. 

There are many measures in the literature to compute the similarity
between these two sets. Among them, the Jaccard ratio is commonly
used. The Jaccard ratio is defined as
\begin{equation}
\label{eq:jaccard}
\Jaccard{\SA}{\SB} = \frac{|\SA\cap\SB|}{|\SA\cup\SB|},
\end{equation}
which maps to $[0,1]$ ---i.e., $1$ if the sets are identical and $0$ if the
sets have no common elements. 

\mypar{Example} Given $\SA=\{a,b,d\}$ and $\SB = \{b,e,f\}$, we have
$\SA\cup\SB = \{a,b,d,e,f\}$ and $\SA\cap\SB = \{b\}$ and thus
$\Jaccard{\SA}{\SB} = \frac{1}{5}=0.2$. 

\subsection{List Similarity}
\label{sec:list-similarity}

As in the measures for comparing sets, there are many measures in the
literature to compute the similarity between two lists. Among them,
Spearman's footrule and Kendall's tau are commonly used. In this
paper, we generalize these measures to include weights and also to
work for partial lists as well as permutations. By also proving the
equivalence of these two measures, we justify our choice of Spearman's
footrule for our list comparison measure.

\subsubsection{Rank Assignment}
\label{sec:rank-def}

Given two lists $\A$ and $\B$, define $\A^c=\A-\A\cap\B$ and
$\B^c=\B-\A\cap\B$ and keep the relative order of the remaining
elements in $\A^c$ and $\B^c$ the same as they are in the original lists
$\A$ and $\B$, respectively. Note that $\A^c$ and $\B^c$ bring forth any
information only when $\A$ and $\B$ are partial lists, because they
are the empty set otherwise (i.e., $\A$ and $\B$ are permutations).

If $\A$ and $\B$ are permutations of length $n$, the rank of an
element $i$ is well defined and equal to $\A(i)$ and $\B(i)$. If these
lists are partial lists, the rank of an element is determined as
follows: If an element $i$ is in $\A$ but missing from $\B$, then let
$\B(i)=n+\A^c(i)-1$; that is, it is like we append the missing items at
the end of the list such as to minimize their displacement.
Similarly, if an element $i$ is in $\B$ but missing from $\A$, then
let $\A(i)=n+\B^c(i)-1$. Now the rank function $\A()$ and $\B()$ infer
two lists that are the permutation of each other. Note that if the
lists are of different lengths, we can always restate the definition
so that if an element $i$ is in $\B$ but missing from $\A$, then let
$\A(i)=|\A|+\B^c(i)-1$. Independently, the resulting lists are
permutations, thus with the same length. 

Of course, this rank extension is arbitrary and relative to the pair
of lists. In fact, we extend the rank of an element that does not
exist in a list (unknown rank) using its rank from another list
(partial known rank). This provides an optimistic ordering that should
bias the permutation-based correlation metrics towards positive
correlation. This way to infer not known rankings is similar/common
for comparing top-k lists \cite{FaKuSi03}. Notice also that we
increased the list size; as a function of the increase, any type of
list increases, we may have made the most common correlation measures
less sensitive.

\mypar{Example} Given $\A=(a,b,d)$ and $\B = (b,e,f)$, we have $\A' =
(a,b,d,e,f)$ and $\B' = (b,e,f,a,d)$; that is, the extended
lists. Now, without loss of generality, we can substitute the letters
to numbers ---i.e., ranks. We take $\A'$ as reference or original
permutation: $\A' = (a,b,c,e,f) \sim (1,2,3,4,5)$ and thus we can
rewrite $\B' = (b,e,f,a,d)$ as $(2,4,5,1,3)$. All measures introduced
in this paper are symmetric, thus the result is independent of whether
we take $\A'$ or $\B'$ as starting point permutation.

\subsubsection{Weighted Spearman's Footrule}
\label{sec:footrule}

The weighted Spearman's footrule \cite{Spearman1906,DiGr77} for
partial lists of length $n$ is defined as
\begin{equation}
\label{eq:footrule}
S_w(\A,\B) = \sum_{i\in\A\cup\B} w(i) |\A(i) - \B(i)| .
\end{equation}
where $w(i)$ returns a positive number as the weight of the element
$i$ and the ranks are defined as in \S~\ref{sec:rank-def}.

The measure $S_w$ can be normalized to the interval of $[-1,1]$ as
\begin{equation}
\label{eq:norm-s-w}
s_w(\A,\B) = 1 - \frac
{2 S_w(\A,\B)}
{\sum_{i\in\A\cup\B} w(i) |(i) - (n - i + 1)|} 
\end{equation}
where the denominator reaches its maximum when both lists are sorted
but in opposite orders.

Both of these equations are valid if the input lists are permutations.

\mypar{Example} Given $\A=(a,b,d)$ and $\B = (b,e,f)$, we have $\A' =
(a,b,d,e,f) \sim (1,2,3,4,5)$ and $\B' = (b,e,f,a,d) \sim (2,4,5,1,3)$
(i.e., we transformed the lists into permutations as we described in
the previous example). Then 
\begin{equation}
  \begin{split}
    S_w =& w(1)|1-4| + w(2)|2-1| +w(3)|3-5|  \\
         & +w(4)|4-2|+ w(5)|5-3|  \\
       = & 10w
  \end{split}
\end{equation}
if we consider $w(i)$ constant $w$ and the normalized 
\[
s_w = 1-2\frac{w*10}{w*12} = -0.66
\]
As we can see the denominator grows as $n^2$

\subsubsection{Weighted Kendall's Tau}
\label{sec:kendall}
In context, the unweighted Kendall's Tau is the number of swaps we
would perform during the bubble sort in such a way to reduce one
permutation to the other. As we described the ranks of the
extended lists (Section \ref{sec:rank-def}), we can always assume that
the first list $\A$ is the identity (increasing from 1 to $n$), and
what we need to compute is the number of swaps to sort the permutation
$\B$ back to the identity permutation (increasing). Here, a weight
will be associated to each swap.

The weighted Kendall's tau \cite{Kendall1938,Sievers1978} for partial
lists of length $n$ is defined as
\begin{equation}
  \label{eq:kendall}
  K_w(\A=\iota,\B)=\sum_{1\leq i<j\leq n} \frac{w(i)+w(j)}{2}[\B(i)>\B(j)]
\end{equation}
where $[x]$ is equal to 1 if the condition $x$ is true and 0
otherwise; also, we identify the permutation $1,2,\dots,n$ simply as
$\iota$. In practice, if we would like to sort in increasing order the
permutation $\B$ using a bubble sort algorithm, then
$K_w(\A=\iota,\B)$ is the cost of each swap. 

The measure $K_w$ can be normalized to the interval of $[-1,1]$ as
\begin{equation}
\label{eq:norm-k-w}
k_{w} = 1- \frac 
{2K_w(\A, \B)}
{\sum_{\{i,j\in\A\cup\B:i<j\}} \frac{w(i) +
    w(j)}{2}} 
\end{equation}
where the value of the denominator is exactly the maximum value that
the numerator can reach: when both lists are sorted but in opposite
orders.

Note that both these equations are computed over all $i$ and $j$ in
$\A\cup\B$ such that $i<j$. They are also valid if the input lists are
permutations.

It is important to note that the weighted version of Kendall's tau can
be defined in different ways (e.g., see \cite{Sh98,ShBaTs00}, the
weights are multiplied as $w(i)*w(j)$) rather than added. The reason
for our definition is to preserve the equivalence between these two
measures, as we prove in a different work \cite{DasdanD2011}.

\mypar{Example} Given $\A=(a,b,d)$ and $\B = (b,e,f)$, we have $\A' =
(a,b,d,e,f) \sim (1,2,3,4,5)$ and $\B' = (b,e,f,a,d) \sim
(2,4,5,1,3)$. Then
\begin{equation}
  K_w = 5w \nonumber
\end{equation}
if we consider $w(i)$ constant $w$ and the normalized 
\[
k_w = 1- \frac{2*5}{5*4/2} = 0
\]
Notice that $K_w\leq S_w \leq 2K_w$ because $5w\leq 10w\leq 10w$. We
show this is true in general \cite{DasdanD2011}.


\begin{figure*}
\centering \includegraphics[width=0.49\linewidth]{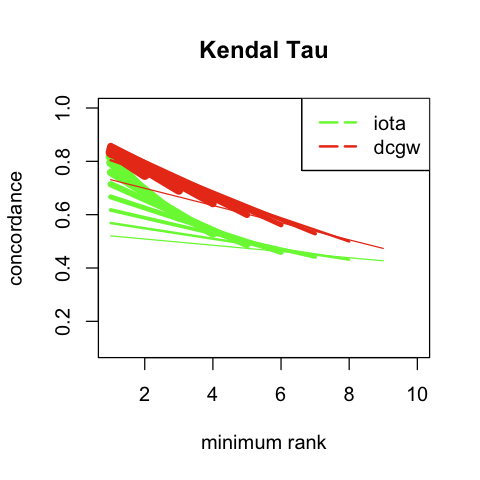}
\includegraphics[width=0.49\linewidth]{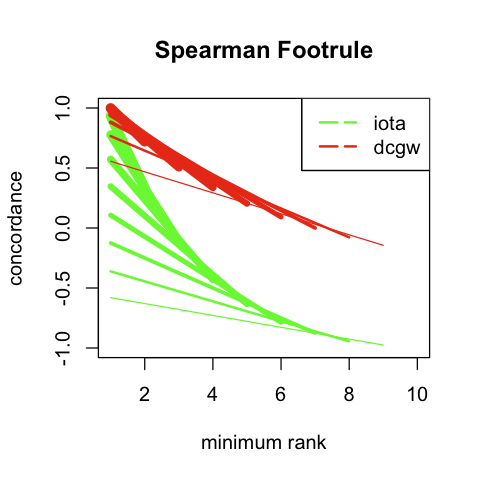}
\\ \caption{In red, we show the list
  concordance measures using weights and in green without weights. The
  width of the lines represents the number of common items in the
  lists: the thinnest lines represent lists with only one item in
  common, the thickest 10 (one point). Consider the thinnest lines:
  Spearman's footrule has the largest difference with weights and
  without; when the two lists have only the first item in common, the
  minimum rank is 1, the weighted measure is about 0.62 and the
  unweighted is 0.18. Both measures decreases when we choose as common
  element the second item towards the ninth element. The weighted
  measures are more sensitive for sparse lists and with high
  correlation in the high ranks.}
\label{fig:weighted-vs-not}
\includegraphics[width=0.49\linewidth]{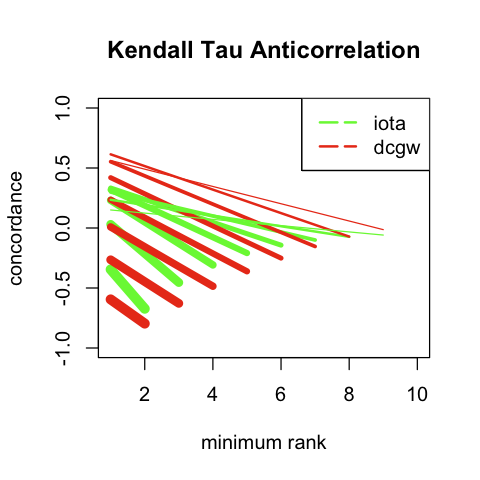}
\includegraphics[width=0.49\linewidth]{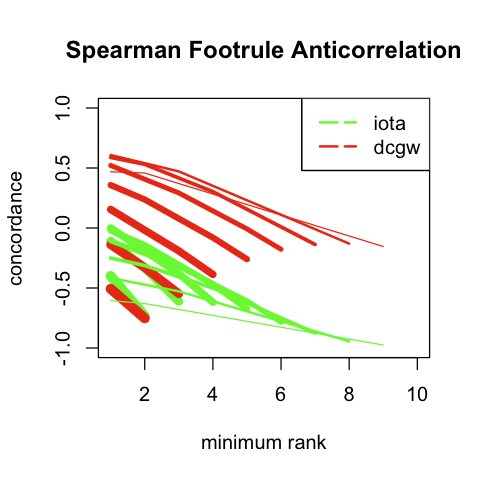}
\\ \caption{ In contrast with Figure
  \ref{fig:weighted-vs-not}, the weighted measure is less sensitive in
  capturing anti correlation especially for sparse lists (less than 4
  common URLs) and for anti-correlation in the high ranks. Notice that
  the unweighted Spearman's footrule finds our lists anti correlated
  (thinnest lines and values close to -1), instead Kendal tau suggests
  no correlation (values close to 0).}
\label{fig:weighted-vs-not-anti}
\end{figure*}

\subsubsection{The weighting function $w()$} 
In this section, we show preliminary evidence that the choice of $w()$
will require a supervised approach and thus beyond the scope of this
paper. Hence, in this paper, we will choose the weighting function
$w()=1$.

We address in this section two questions: First, will a weighted
measure be useful for sparse lists comparisons (for search engines
results)?  Second, what is the choice of the weighting function?
Weighted measures are useful because they provide a way to measure the
importance of common items in the results lists so that to complete
the missing information about the lists we compare. For example, if we
have a query, two search engines results (10 URLs each list), and we
find out that there are only four common results; we can estimate a
measure of disarray/concordance if we can assign a heavier weight for
higher URLs (on the list top).

Here we choose two weighting functions that we identify as {\em dcgw}
and {\em iota}. We have $iota(i)=1$ for every $i$: that is, all list
URLs are equally important. Instead, we have $dcgw(i)=
\frac{log_{10}(1+i)}{2^i}$, which is inspired by the discounted
cumulative gain (DCG) measure; that is, we can imagine that the tenth
URL in the result list is about $2^9$ less important than
---relatively speaking--- the first one.

We created this test: we take two lists $a=(1,2,..,10)$ and
$b=(11,12,..,20)$, and we consider these lists composed of the simbols
'1' throught '20' and thus with nothing in common. We start creating
lists with increasing common intersection: $a$ and $b'$ with one
common item: $b'=(1,12,..,20)$, $b'=(11,2,..,20)$, till $b'=(11,12,
.., 10)$, then with two consecutive items $b'=(1,2,..,20)$,
$b'=(11,2,3,..,20)$, till $b'=(11,12, ..,9, 10)$ and eventually with
10 common consecutive items $b'=a$.

In Figure \ref{fig:weighted-vs-not}, we present the concordance measure
results using Spearman's footrule and Kendal Tau. In red, we show the
list concordance measures using weights and in green without
weights. The width of the lines represents the number of common items
in the lists: the thinnest lines represent lists with only one item in
common, the thickest 10 (one point). Consider the thinnest lines:
Spearman's footrule has the largest difference with weights and
without; when the two lists have only the first item in common, the
minimum rank is 1, the weighted measure is about 0.62 and the
unweighted is 0.18. Both measures decreases when we choose as common
element the second item towards the ninth element. The weighted
measures are more sensitive for sparse lists and with high correlation
in the high ranks.
In Figure \ref{fig:weighted-vs-not-anti}, we show the same analysis
but instead of creating similar lists, we create anti-correlated
similar lists.  The weighted measures are less sensitive in capturing
anti correlation. Even thought this is an example where the same
weighting function achieves contrasting and opposite results, it shows
a case where the function choice must rely on the context for which
the function is applied for. In this work, we are actually interested
in finding a measure that can capture both properties.

At this time, we believe that the solution must rely on a supervised
method where a third party (a crowd base similarity measures) or a
feed back system can be deployed to tune the weighting function
knowing the context. In a different work, we prove that the weighed
Spearman's Footrule and Kendall Tau for partial lists (as described
here) respect the Diaconis-Graham inequality, thus they are equivalent
as discriminative power and we can choose either one (footrule because
of its computational simplicity) \cite{DasdanD2011}.

\subsection{Distribution Similarity}
\label{sec:dist-similarity}

As in the measures for comparing sets and lists, there are many
measures in the literature to compute the similarity between
distributions ---i.e., stochastic distances. For example, a document
can be represented as a word--count histogram (which can be normalized
naturally to a distribution), and this idea can be easily extended to
a set of documents. Among distribution measures, we present and use
here the $\phi$ measure from \cite{KiferBG2004}, which is identified
in \cite{DAlbertoD09} as one of the best performing measures. The
$\phi$ measure extends the well-known Kolmogorov-Smirnov measure and
is defined as
\begin{equation}
\label{eq:phi}
\phi(\FA,\FB) =
\max_{i}\frac{|\FA(i)-\FB(i)|}{\sqrt{\min(\frac{\FA(i)+\FB(i)}{2},
1-\frac{\FA(i)+\FB(i)}{2})}}
\end{equation}
where $\FA$ and $\FB$ are the cumulative distribution functions and
$\FA(i)$ and $\FB(i)$ are the values for the element $i$ from these
functions. This measure is symmetric and its value ranges in $[0,2]$,
where the result is zero when two input distributions are
identical. In practice, we can use stochastic distances to compare the contents
of search engines results. 

We can also determine whether or not two documents are duplicate by
using a set of these stochastic measures and use their confidence
levels to flag equality/difference by a consensus based approach (see
Section \ref{sec:reduction-and-normalization} and \cite{DAlbertoD09});
that is, if the measure majority suggests equivalence, we consider the
document duplicates, otherwise they are not duplicate.  Of course,
stochastic measures will compare distributions, so we are really
saying that two distributions are similar, we infer that they bring
forth the same information, then we deduce that the documents can be
considered duplicates or having similar contents.

\mypar{Example} Consider two documents as a sequence of letters $\A =
(a,b,e,a,e)$ and $\B = (h,a,e,a,)$, the histogram representation will
be $h_\A = (a=2/5,b=1/5,e=2/5)$ and $h_\B = (a=2/4,e=1/4,h=1/4)$, a
possible cumulative distribution extension is $F_\A =
(a=2/5,b=3/5,e=5/5,h=5/5)$ and $F_\B = (a=2/4,b=2/4,e=3/4,h=4/4)$,
thus $\phi(\FA,\FB)= 0.7$, they are different.

\section{Application of the Theory}
\label{sec:practice}
We next detail how we applied the theory of similarity to compute the
similarity between search results from different search engines. For
each case, we took only (up to) the top 10 URLs. Of course, we could
extend the investigation to any number of URLs; however, as almost all
users pay attention to the first page only and because we do not to
try to fuse the list into a single one, the results here presented are
more representative than say the collection of the first 100
URLs. \footnote{Notice that as the result list gets longer, the more
  the correlation measures such as footrule is less sensitive and the
  confidence level drops drastically artificially creating a scenario
  where we cannot say anything about correlation either way.}

\subsection{Search Results as Sets}
Search results $\A$ and $\B$ from two search engines for the same
query can be represented as either two sets of URLs or two sets of
contents, which are the terms extracted from the landing pages or
documents. 

As sets, the rank of any URL in the original search results was
ignored in the final representation. We kept a unique copy of any
element in the final lists: the duplication test was done using the
shingling technique~\cite{Broder1997,He06} over the landing page
contents of the URLs (see \S~\ref{sec:similarity-duplicates} for
details). Thus for a set of URLs, the duplicate detection is used to
normalized the URLs and thus the lists; this is necessary, because
different search engines may use different policy for the canonical
representation of a URL. As a set of contents, no URL normalization is
necessary and simply the contents union of the landing pages is used
instead.

We used the Jaccard ratio to compare the resulting sets. In the
sequel, we use the notation $J_{url,n}$ and $J_{term,n}$ to denote the
Jaccard ratio between the sets of $n$ URLs and the contents of the
corresponding landing pages, respectively. We provide a detailed
description of the use of the Jaccard ratio for the contents in the
following Section \ref{sec:similarity-duplicates}.

\subsection{Search Results as Lists}
Search results $\A$ and $\B$ from two search engines for the same
query can be represented as two lists of URLs. We kept a unique copy
of each URL in the final lists (see previous section for the duplicate
policy). We showed only Spearman's footrule to compare the resulting
URL sets because of the equivalence justification. In the sequel, we
use the notation $s_{url,n}$ to denote the normalized version of
Spearman's footrule between two lists of $n$ URLs.

\subsection{Search Results as Distributions}
\label{sec:distributions}
Search results $\A$ and $\B$ from two search engines for the same
query can be represented as two distributions of term frequencies. We
downloaded the landing page contents of the search result URLs. We
extracted terms and their frequencies in each document. To give weight
to top search results, we created distributions from the top-$n$
search results for different values of $n$. We used the $\phi$ measure
to compare the resulting distributions. In the sequel, we use the
notation $\phi_{term,n}$ to denote the $\phi$ measure between two
distributions of landing page contents from two sets of $n$ URLs.


%
%
%
%

\section{Experimental Methodology}
\label{sec:method}

Using a fully automated process, we have been collecting and recording
the performance of two major search engines for about two months for a
total of up to 1,000 queries per day for about 20 countries (50
queries a day per country) a few may address the country as a region
but as we still show in the following the terminology is completely
immaterial. For brevity, we will focus on 4 representative countries
in the sequel.  In this section, we present how we chose our queries,
how we extracted search results and their landing pages, and how we
computed similarity.

\subsection{Sampling Queries}
\label{sec:sampling}

Users submit a stream of queries every day. These queries are easily
classified geographically based on the country of the origin where the
query was submitted; for example, United States (US), Japan (JP),
France (FR), and Taiwan (TW). For each country, a uniformly random
query subset sample is selected out of the entire query stream
daily. This original sample had one million queries a day and is used
by multiple internal customers (within Yahoo!). To make the scale of
our experimentation manageable, we performed another uniformly random
selection of 1,000 queries (about 50 per country) out of this
sample. To reduce the sampling error, we used the stratified sampling
technique with three strata of highly frequent, frequent, and
infrequent queries and sampled from each stratum with equal
probability. So our sample set contains frequent queries as well as
tail queries in equal amount; the sampling is time sensitive so that
the same query is very unlikely to be chosen, day after day. Thus
overall, we have a balanced set and frequent queries should not bias
our results, and so the tail queries. \footnote{Frequent queries are
  usually the queries where all engines do well because they get
  trained in time, and our results will show a consistent divergence.}

So we do not classify the queries and we do not use any taxonomy or
classification of the queries such as navigational,
commercial. Unfortunately, most of this classifications are based on
explicit human judgments (editors) or user-behavior feedback
measures. These are beyond the scope of this work (unsupervised) but
of course could be applied to the methodology. At the same time, we
will show a comparison for representative queries that are used for
relevance measures and where this classification is in place; in this
scenario, we will show that our main message does not change, there is
a divergence in the results-list contents but not necessarily in the
quality of the user satisfaction.

Let us express one last note about query selection: similar queries,
which differ very little, can be selected in this process. The fact
the sampling is done during a long period of time and for different
countries (different needs) and further stratified should alleviate
any bias towards these similar queries. Interesting enough, out
techniques can be used in practice just to find similar queries by
looking at the search result lists.

\subsection{Scraping Results}

Each day and for each country, we repeated the following process: we
submitted the queries from our daily query sample to a number of major
search engines (in terms of the market share) and scraped the returned
results. A query coming from France is sent to the search engines so
that to reproduce the results as a French user would see from his
laptop in Paris. So each engine can provide a custom experience for
the same query in different markets.

For each returned URL, we downloaded the landing page contents using
our production crawler. \footnote{If we do not have it, and the site
  allows us crawling, we actually fetch the document.} Finally,
we compared the similarity between every pair of search engines using
all the similarity measures discussed in \S~\ref{sec:practice}.

\subsection{Deciding Duplicate Contents} 
\label{sec:similarity-duplicates}

For reasons of practicality, we performed our content-based similarity
computation over shingles~\cite{Broder1997,He06} instead of raw
terms. In other words, in Eq.~\ref{eq:jaccard}, the sets $\SA$ and
$\SB$ contained shingles rather than terms. We used 1,000 shingles
with 10 consecutive terms per shingle. So given two items $\A(i)$ and
$\B(j)$, with the term $J_{term,1}(\A(i),\B(j))$, in the duplicate
detection context, we mean $\Jaccard{{\cal S}_{1000}(\A(i))}{{\cal
    S}_{1000}(\B(j))}$ where ${\cal S}_{1000}(\B(j))$ is the set of
the first 1000 shingles of document $\B(i)$ and thus
\[ J_{term,n}(\A,\B) = J_{term,1}(\cup_i^n {\cal
  S}_{1000}(\A(i)),\cup_j^n {\cal S}_{1000}(\B(j))). \]

Given this measure, we regarded two sets as duplicate in contents if
their Jaccard ratio was above 0.5. This threshold choice is based on
our previous experience with duplicate detection techniques. An
intuitive suggestion about this threshold is that, when a document has
more that 60\% of the contents ---as 10 word long sentences and
considering 1000 of these--- are common to another document, then the
probability to have two different documents is ridiculously small
especially for large documents.

\mypar{Example} Consider a document as a sequence of letters $\A =
(a,b,c,a,b,c)$ and consider a window of size 3 letter (shingle). We
obtain four shingles $s_0=(a,b,c)$, $s_1=(b,c,a)$, $s_2=(c,a,b)$ and
$s_4=(a,b,c)$. In general, if the document has $n$ words and the
shingle window is of size $m$, we have up to $n - m$
shingles. However, $s_0 = s_4$ and we do not consider the multiplicity
of a shingle and the document is summarized by only three shingles
$s_0, s_1$, and $s_2$. Thus, we have $\SA = \{s_0, s_1,s_2\}$. In
practice, the shingles are encoded by a unique integer and we have a
set of integers (letters if you will) and then we can apply the
Jaccard ratio. If we take $\B = (c,b,a,c,b,a)$, which is the inverse
of $\A$, we have four shingles but will keep only three:
$t_0=(c,b,a)$, $t_1=(b,a,c)$ and $t_2=(a,c,b)$. $J_{term,1}(\A,\B) =
\frac{0}{6}$. The two documents are not duplicate.

For deciding two documents as duplicate when using distributions, we
computed their similarity using the following 10 distribution
similarity measures from \cite{DAlbertoD09}: $\phi$, $\Xi$,
Kolmogorov-Smirnov, Kullback-Leibler, Jensen-Shannon, $\chi^2$,
Hellinger, Carmer-von Mises, Euclid, and Canberra. If more than 4 out
of these 10 flagged two documents as duplicate with a statistical
significance level of 5\%, we considered the input documents as
duplicate.  So given two items $\A(i)$ and $\B(j)$, with the term
$\DELTA{\A(i)}{\B(j)}$, in the duplicate detection context, we
actually mean the comparison above with 10 stochastic measures and
using distributions so that $\A(i)$ is a duplicate of $\B(j)$ if and
only if $\DELTA{\A(i)}{\B(j)}=1$, and as not duplicates iff
$\DELTA{\A(i)}{\B(j)}<1$ (and thus $\DELTA{\A}{\B} \equiv
\DELTA{\cup_i \A(i)}{\cup_j \B(j)}$). Notice we store each measure
separately and thus we can apply each measure to a single pair of
documents as well as to any subset of the result list.

\mypar{Example} Consider two documents as a sequence of letters $\A =
(a,b,c,a,b,c)$ and $\B = (c,b,a,c,b,a)$ as before. These documents
will have histogram $h_\A= (a=1/3, b=1/3, c=1/3)$ and $h_\B =h_\A$. As
expected, the two documents will be considered duplicate. These
measures if applied for duplicate detection are looser than the ones
based on shingles: Using shingles, we may consider documents as not
duplicate but they are, using distributions we may consider documents
as duplicate but they are not.


\subsubsection{Reduction to  Spearman's Footrule}
\label{sec:reduction-and-normalization}

Assume we have two lists of URLs and we want to compare their
correlation. Since these lists are coming from different engines we
cannot assume that the same documents have the same URL. We need to
bind the document to a single URL or name and then we can perform any
list based comparison. We propose to use the similarity functions in
such a way to perform the unique document-URL binding. As result of
this URL normalization we are able to enlarge, when possible, the
Jaccard ratio of the lists and making the correlation better suited.
Here we explain how we do it.

Take the two lists $\A$ and $\B$, we start with $\A$ and we are going
to rewrite it to $\tilde{\A}=(\Ai{0})$ ---i.e., the list containing
only the {\bf first} URL or item--- and $\B$ to $\tilde{\B}=()$
---i.e., the empty list.  We use the similarity function
$J_{term}(,)$ ---shingle based comparison as in Equation
\ref{eq:jaccard}--- and $\DELTA{}{}$ ---Histogram--CDF based
comparison as in Equation \ref{eq:phi}.  Here, we present our
URL-normalization algorithm for two lists:

\mypar{$(\tilde{\A}_\omega,\tilde{\B}_\omega) =$ {\bf
    Normalization}$(\A,\B)$}
\begin{enumerate}
\item For every $i>1$ and $\Ai{i} \in \A$ (in the order of the
  original list, from the highest rank to the lowest)
  \begin{enumerate}
  \item $Image(\Ai{i})$ is the set $\{ v \in \A \text{ so that }
    J_{term,1}(v,\Ai{i})\geq 0.5 \}$.
  \item $C = \tilde{\A}\cap Image(\Ai{i})$ is the set of duplicates we
    have already seen.
    
  \item if $|C|>0$ then append the first element in $\tilde{\A}$ that
    is in $C$ to $\tilde{\A}$
  
  \item else append $\Ai{i}$ to  $\tilde{\A}$ 
  \end{enumerate}
\item  For every $i\geq 1$ and $\Bi{i} \in \B$
  \begin{enumerate}
  \item $Image(\Bi{i})$ is the set $\{ v \in \B \text{ if }
    J_{term,1}(v,\Bi{i})\geq 0.5 \text { or } v \in \A \text{ if }
    \DELTA{v}{\Bi{i}} =1 \}$.
  \item $C = \tilde{\A}\cap \tilde{\B}\cap Image(\Bi{i})$ is the set of duplicates we
    have already seen.
  \item if $|C|>0$ then
    \begin{enumerate}
    \item append the first element in $\tilde{\A}$ that is in $C$ to
      $\tilde{\B}$, if any (priority to the first list)
    \item append the first element in $\tilde{\B}$ that is in $C$ to $\tilde{\B}$, otherwise
     \end{enumerate}
  \item else append $\Bi{i}$ to  $\tilde{\B}$ 
  \end{enumerate}
\end{enumerate}

As a result, duplicate items are relabeled using a single name. Across
different lists, this is an efficient URL normalization (independent
of the search engines) and it increases the lists intersection
naturally. A side effect, of this lists normalization, is that we are
going to flag out duplicates within the same list (and also across
lists and especially for the second list).  Then, we need to penalize
any search engine that introduce duplicates. We post process the lists
so that any subsequent duplicate within a list will be substitute with
a empty item $\omega$, which will be taking the ranking position but
it will not be used for any comparison. In Section
\ref{sec:low-overlap} and in particular in
Fig. \ref{fig:Jaccard-and-query}, we will show that the way we perform
the URL normalization across lists has very little effect and thus the
little overlap it is not due to the way we perform the normalization.



\mypar{Possible extension} We could use the
$\Jaccard{\tilde{\sigma}_\omega}{\tilde{\pi}_\omega}$ to provide a
normalizing factor for the normalized measure $\tilde{s}_w(\A, \B)$ so
that to extend the range of the measure to original interval [-1,1]
and thus possibly use the footrule distribution function. This is
beyond the scope of this work but a natural extension.

\section{Results on Search Results Similarity}
\label{sec:experimental-results}

\Doublefigure{0.49}{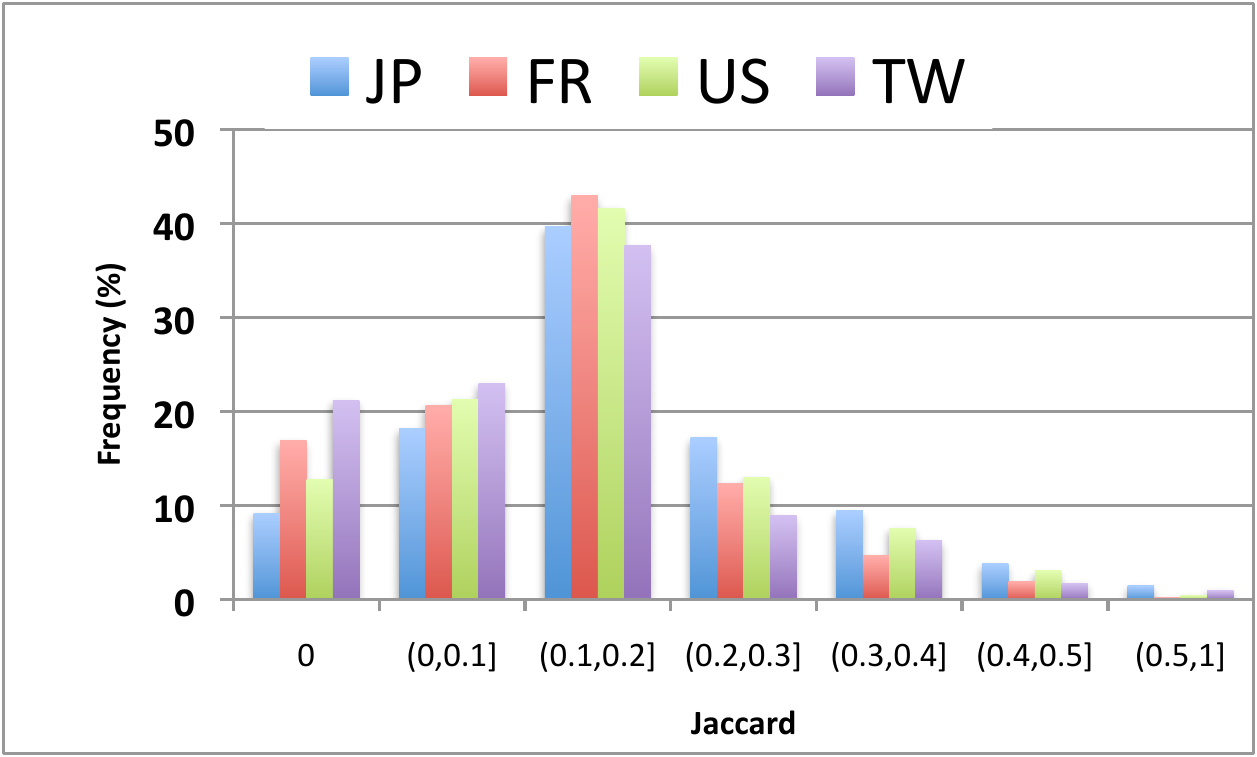}{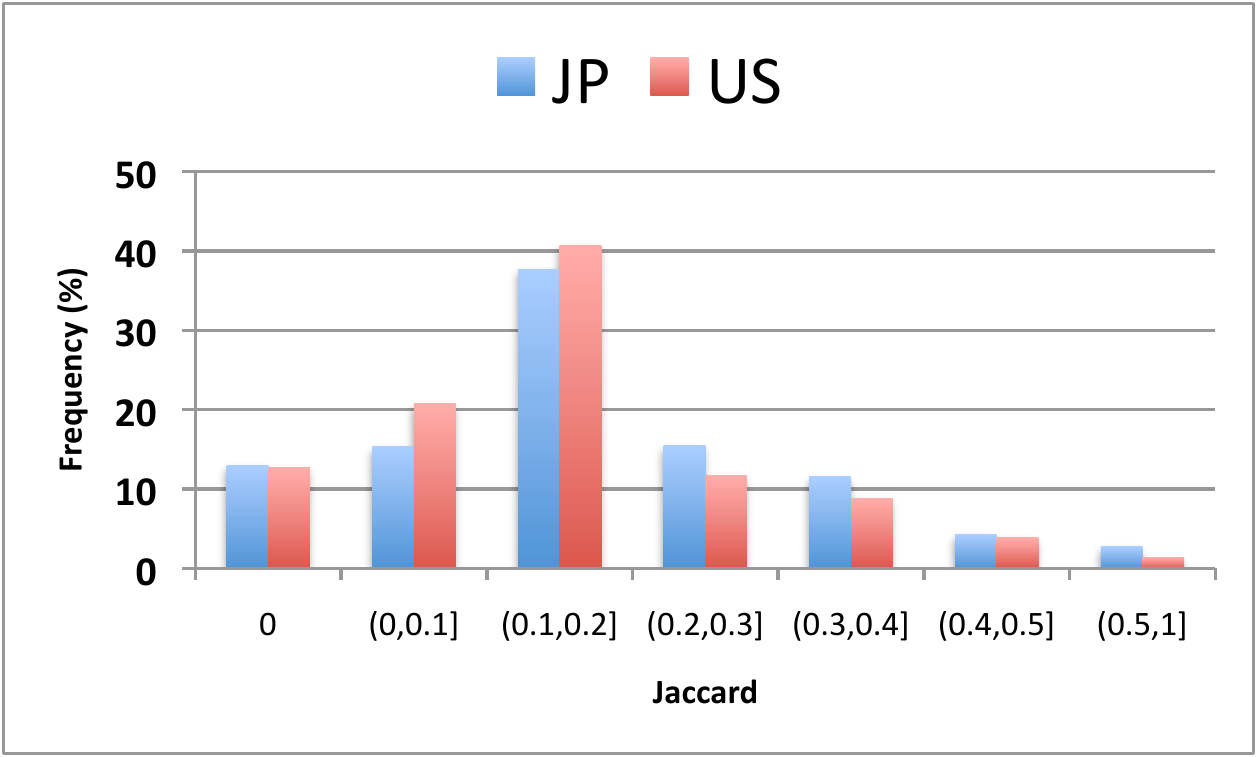}{Number of
  queries and equivalent number of common results expressed as the
  Jaccard ratio of the URLs $J_{URL,10}$: left, duplicate using
  $J_{term,1}(,)$ only (shingles); right, using both $J_{term,1}(,)$ or
  $\DELTA{}{}$ (loose comparison) for only US and JP where we have
  more intersection to start with. The bars are in the same order as
  the legend. Note the way we compare duplicates between lists does
  not change the divergence of the result
  lists overall.}{fig:Jaccard-and-query}

We present our observations on search results similarity in terms of
the evolution of the overlap between search results as well as the
correlation between overlap and quality.

\subsection{Low Overlap}
\label{sec:low-overlap}

For a reliable data point in the past, we refer to \cite{FaKuSi03}. In
this reference, pairwise URL-based similarity of seven search engines
over 750 URLs is computed using a version of Kendall's tau. It shows
that search engines produce quite different results except in the case
of having the same third party provider of crawled contents. Another
confirmation of the low overlap comes from \cite{Bar-IlanML2006} where
the overlap is found to be ``very small''. However, both of these
studies use very few queries for supporting their findings.

\begin{obs}
  For more than 80\% of the queries, the overlap between two sets of search
  results is less than 30\%.
\end{obs}

To support this finding, we present Fig.~\ref{fig:Jaccard-and-query},
where a histogram for the overlap in URLs is given for four
representative markets. The x-axis shows the Jaccard ratio
($J_{url,10}$) as an interval for the URLs common to both lists and
the y-axis shows the frequency of overlap.

In this figure, markets have similar behavior. The highest frequency
bucket across markets is that for about 40\% of the queries the
overlap as measured by the Jaccard ratio fall into the interval $(.1,.2]$
(i.e., between 2 and 3 common URLs within top-10 results). If we add
up the first three buckets, then we get what the observation
claims. Over all markets, less than 5\% of the queries have more than
7 common URLs. 

In Fig.~\ref{fig:Jaccard-and-query}, we also show that the low overlap
is independent of the duplicate detection measure used from a stricter
(on the left) to a looser (on the right). Notice that we performed in
parallel for JP and US the same process (of course with queries chosen
randomly and independently as described in Section \ref{sec:sampling})
only using a different duplicate detection. We chose JP and US because
in this example we have the largest overlap. This shows that the
little overlap is because of the documents in the lists instead of the
way we perform the tests (inherent property of the engines
results). Of course, if we do not apply any duplicate detection, the
overlap will be even lower exaggerating the divergence of the result
lists. Previous works show overlap only for URL-based comparisons, so
we show a larger divergence with a stronger approach for the lists
comparisons. Nonetheless, despite our best effort to bring forth more
common URLs in the lists, the overlap is very limited and decreasing.

If we would apply list-based correlation measures on this set as in
the previous works, we will find little correlation (or un-correlation
for that matter) because there is little overlap not because there is
a real correlation. We will come back to this in Section
\ref{sec:urls-vs-contents}.

%
%

\subsection{Varying Quality: quality vs. overlap}
\label{sec:varying-quality}

The results in this section require a quality definition and
measurement. By the quality of a set of search results for a query, we
mean the relevance of the search results in satisfying the information
need of the user expressed by the query (e.g., see \cite{DasdanTV2009}
for a detailed discussion on relevance).

Among the measures to quantify the relevance, Discounted Cumulative
Gain (DCG)~\cite{JaKe02} seems to be the measure preferred by most
search engines. For a query set $Q$, each with $n$ ranked search
results, DCG is defined as
\begin{equation}
\label{eq:dcg}
DCG_n = \frac{1}{|Q|}\sum_{q\in Q} DCG_n(q) \;\mbox{and}\; DCG_n(q) =
\sum_{r=1}^{n} \frac{g(r)}{d(r)}
\end{equation}
where $g(r)$ is the gain for the document for the URL at rank $r$ and
$d(r)=\lg(1+r)$ is the discounting factor to bias towards top
ranks. Typically, $g(r)=2^{j-1}-1$ where $j$ is equal to 5, 4, 3, 2,
or 1 for the judgments of Perfect, Excellent, Good, Fair or Bad
results, respectively. The judgments are from editors binding the
query intent to the result lists and their document contents.

In this section, we used about 800 queries selected uniformly at
random from user queries submitted to our search engine (some are
identical queries at different times). For each query, we scraped the
top 5 search results for each of the major search engines.

We define a relative measure between two search engines $SE_1$ and
$SE_2$ as
\begin{equation}
r_{DCG_5}(q) = \frac{DCG_5^{SE_1}(q) - DCG_5^{SE_2}(q)}{\max(DCG_5^{SE_1}(q),DCG_5^{SE_2}(q))}
\end{equation}
where $-1\leq r_{DCG_5}(q) \leq 1$ and $DCG_5^{SE_i}$ is the DCG for
$SE_i$ with $i=1,2$ (where we leave the true identity of the search
engine anonymous for obvious reasons).

\doublefigure{0.80}{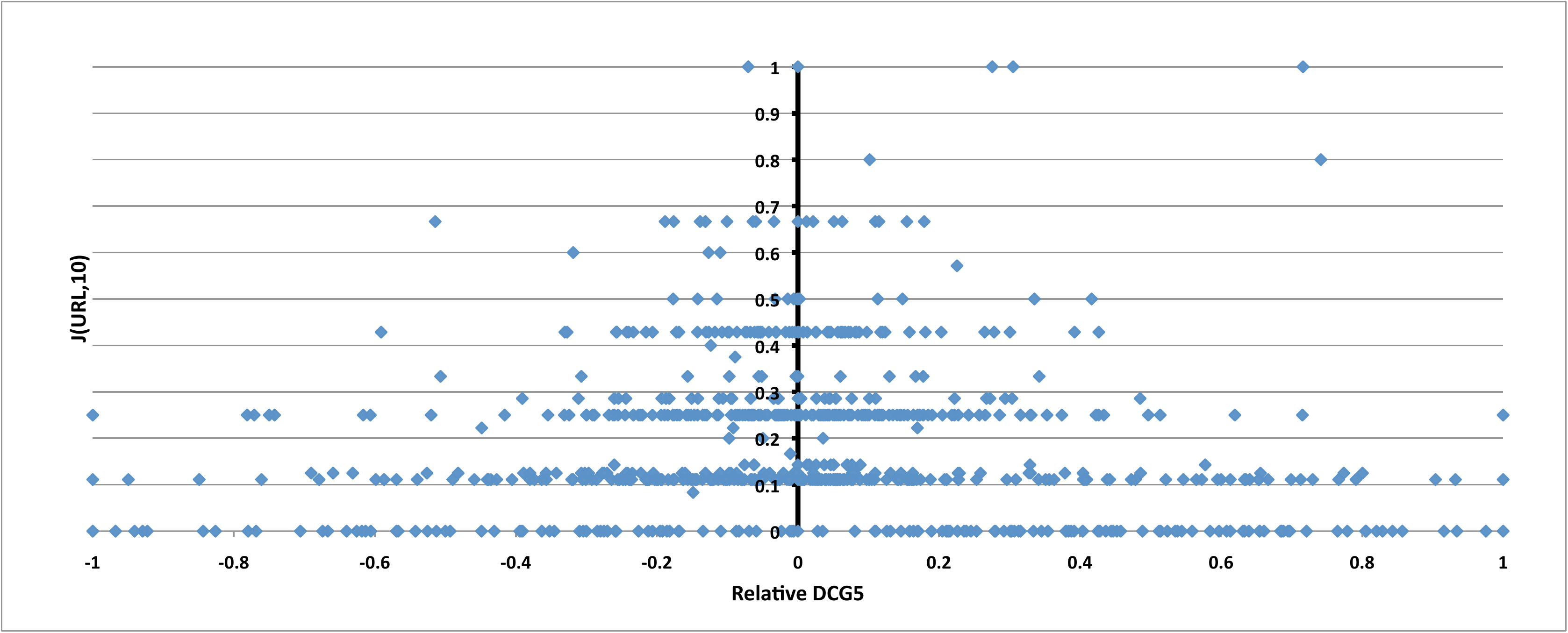}{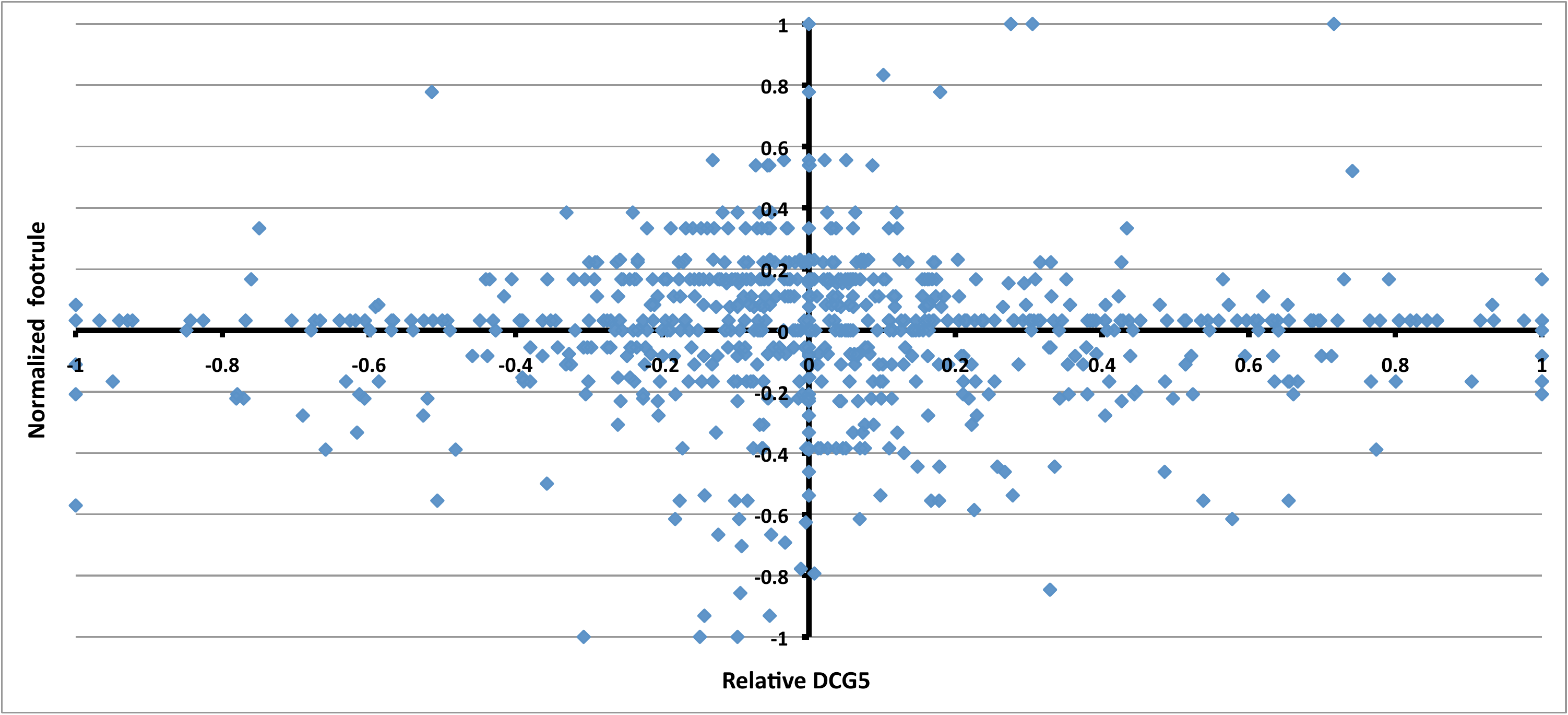}{Top: $J_{url,5}$
  vs. $DCG_5$ with greatly varying DCG when the overlap is low;
  bottom: $s_{url,5}$ vs. $DCG_5$ with greatly varying DCG at no
  correlation. }{fig:DCG-and-lists}

Armed with these definitions and results, we can state the
relationship between quality and overlap. 

\begin{obs}
When the overlap is low between the results of two search engines, the
relative quality between search engines varies widely.
\end{obs}

In Fig.~\ref{fig:DCG-and-lists}, we present a comparison between the
relative $DCG_5$, $s_{url,5}$ (footrule) and $J_{url,5}$ (common
intersection). There is no correlation (not upon intended) between DCG
and footrule. However, we can say that when the common intersection
between the result list is large enough there is no particular
difference of DCG values and thus the search engines seem correlated
and equivalent (i.e., having a large number of common URLs and the
editors graded these URLs with similar scores based on contents and
ranking, we can safely infer that the engines provide the same URLs
and with the same ranking). In other words, low overlap does not
necessarily mean that one of the search engines is consistently better
at the search results quality.

\singlefigure{0.5}{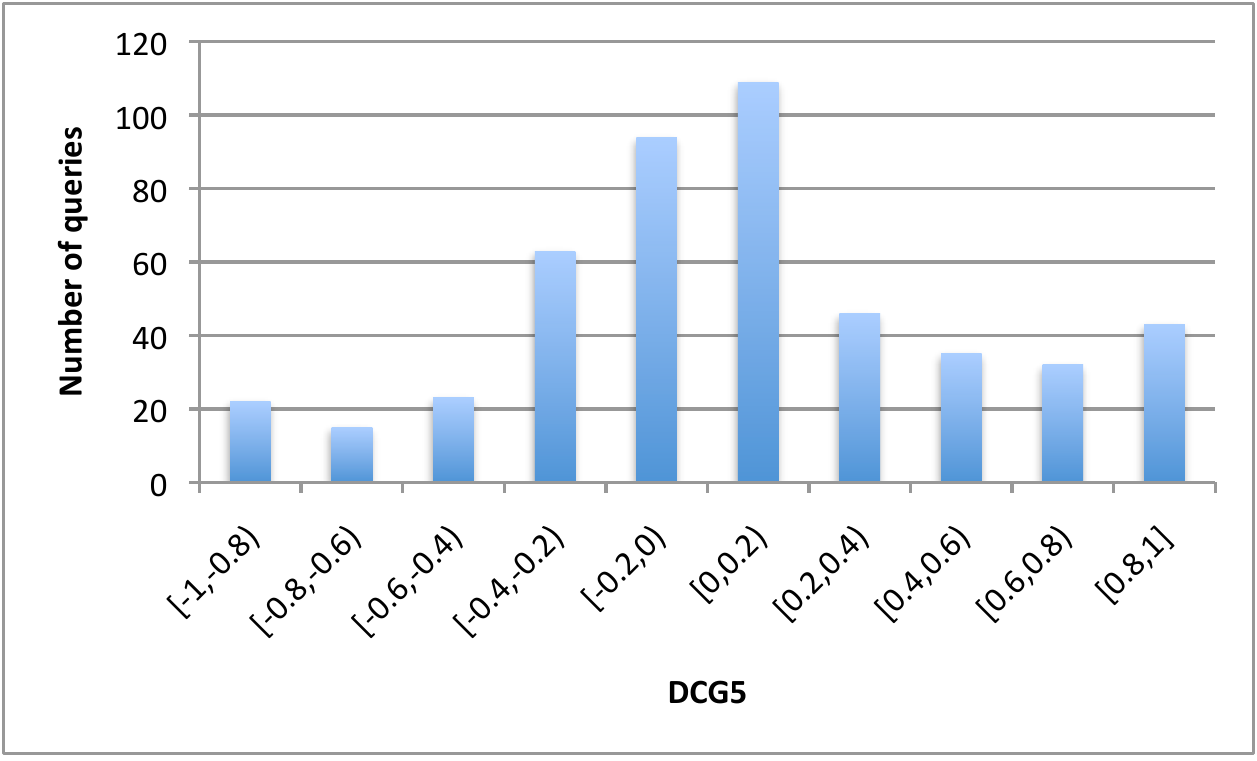}{Distribution of the relative DCG
  when the overlap is low: with Jaccard less than
  0.2.}{fig:hist-dcg-and-jaccard}

To support this finding, we present Fig.~\ref{fig:DCG-and-lists},
where the scatter plot between the DCG and two similarity measures is
given. The x-axis shows the relative DCG measure and the y-axis shows
the Jaccard ratio (top) and normalized footrule (bottom).

We have three observations: First, most of the queries have low
overlap as measured by the Jaccard ratio; second, most of the queries
fall into the narrow interval [-0.2, 0.2] as measured by the
normalized footrule; third, for most of the queries the DCG value is
orthogonal to both measures. One application of list based measures is
the determination of query with little overlap to filter/reduce the
list of queries that really need editorial judgments.

In Fig.~\ref{fig:hist-dcg-and-jaccard}, we provide additional evidence
for our third observation (i.e., for most of the queries the DCG value
is orthogonal to both measures). In this figure, we present the
distribution of the $r_{DCG_5}(q)$ over all queries $q$ such that the
Jaccard ratio is less than 0.2; that is, with low overlap (1 in 5
common results).  The existence of fat tails at both ends of the
distribution implies a large range of values for quality as measured
by the DCG (most likely the quality of the search results does not
{\em come} from the common results). 


\subsection{Results on Similarity Measures}
\label{sec:urls-vs-contents}

We conclude the experimental section by stating our last observation,
which is one of the main motivations for this work.

\begin{obs}
Due to the low overlap between search results, content-based
similarity measures provide more discriminating conclusions than
URL-based similarity measures do.
\end{obs}

We are going to break down the discussion into four parts: the
relationship of content-based and URL-based Jaccard ratios (i.e.,
different ways of measuring overlap), URL- and content-based measures
and normalized Spearman's footrule (i.e., overlap vs. rank
correlation), the effect of contents size on the similarity outcome
(i.e., parameter sensitivity of content-based measures), and the
relationship between content-based measures and relative DCG (i.e.,
overlap vs. quality).


\doublefigure{0.80}{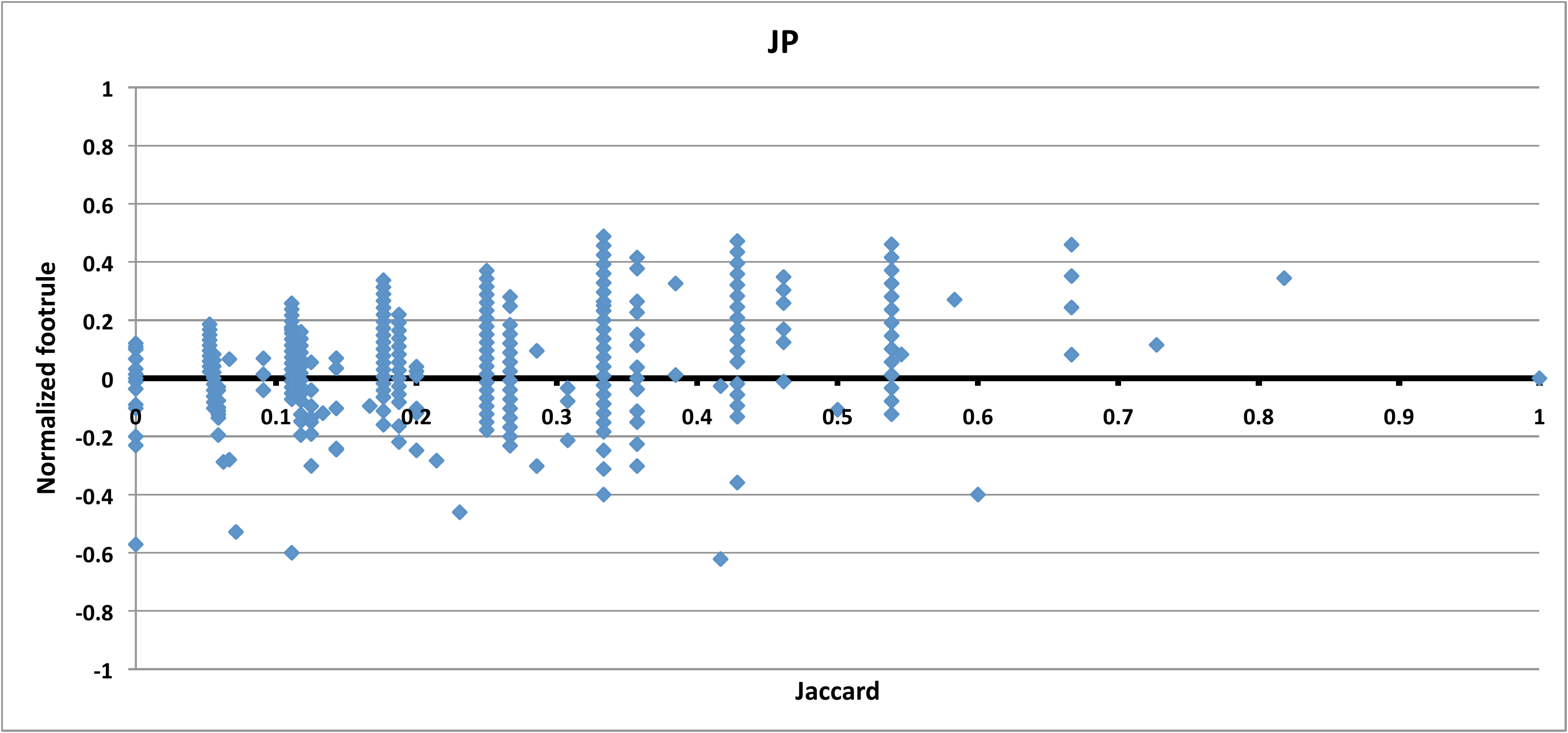}{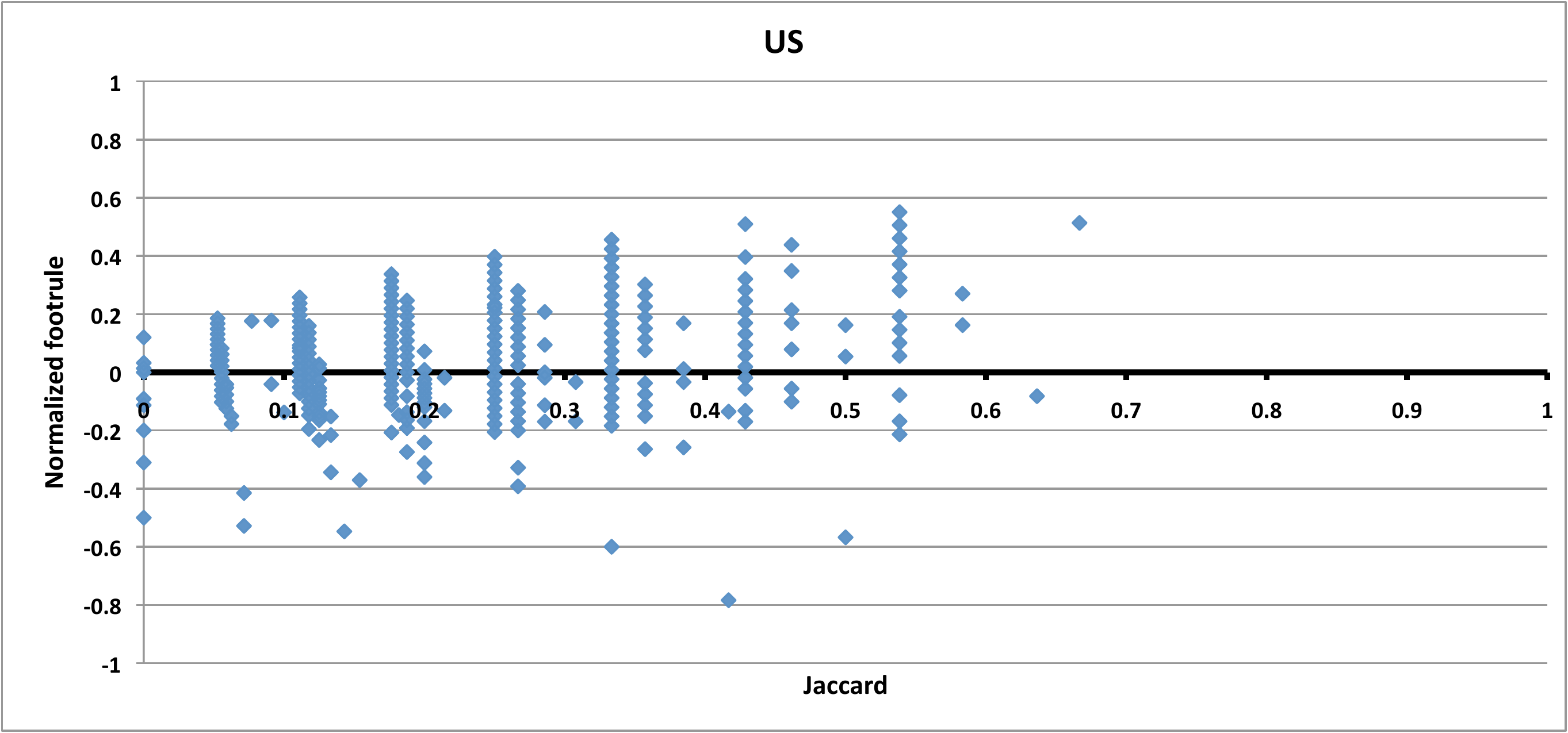}{Comparison
  of the URL-based Jaccard ratio and normalized footrule: $J_{url,10}$
  vs $s_{url,10}$ }{fig:footrule-and-jaccard}

\subsubsection{URL-based Jaccard ratio vs. normalized footrule}

In practice, here we present how weak lists based correlation measures
are and we show in a plain cross product (scatter plot) that little
correlation (or lack of correlation thereof) is because the lists have
really small intersection.

In Fig.~\ref{fig:footrule-and-jaccard}, we show the relationship
between the overlap of the URLs and the normalized footrule for two
markets US and JP. As soon as the overlap of the lists decreases, the
range of the normalized footrule also shrinks. Thus, at low overlap,
list similarity measures will be less meaningful and less
discriminating. Even the power of the URL-based Jaccard ratio
decreases, helping support the need for content-based measures.

If we wanted to use Spearman's foot rule as correlation measure we
would be tempted to assume there is little correlation between the
lists even in the most favorable cases. Actually, we have halved the
range of the measure and thus what we really miss is the confidence in
the measure more than missing a correlation measure. Thus, these
measures have little or no discriminative power. List-based measure
are not suitable for neither automatic nor unsupervised methods.
 
\subsubsection{Content-based vs. URL-based Jaccard ratios}

\doublefigure{0.80}{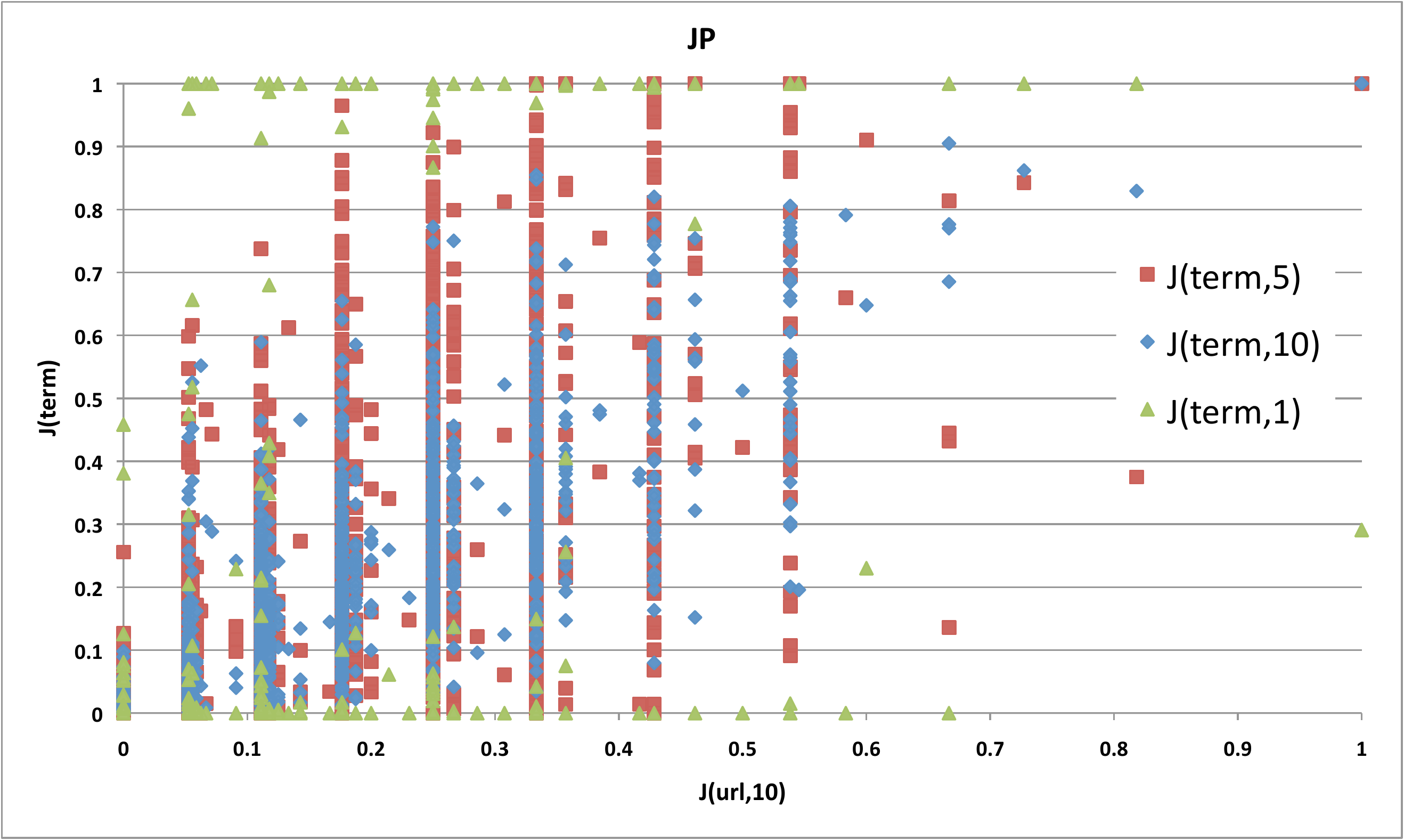}{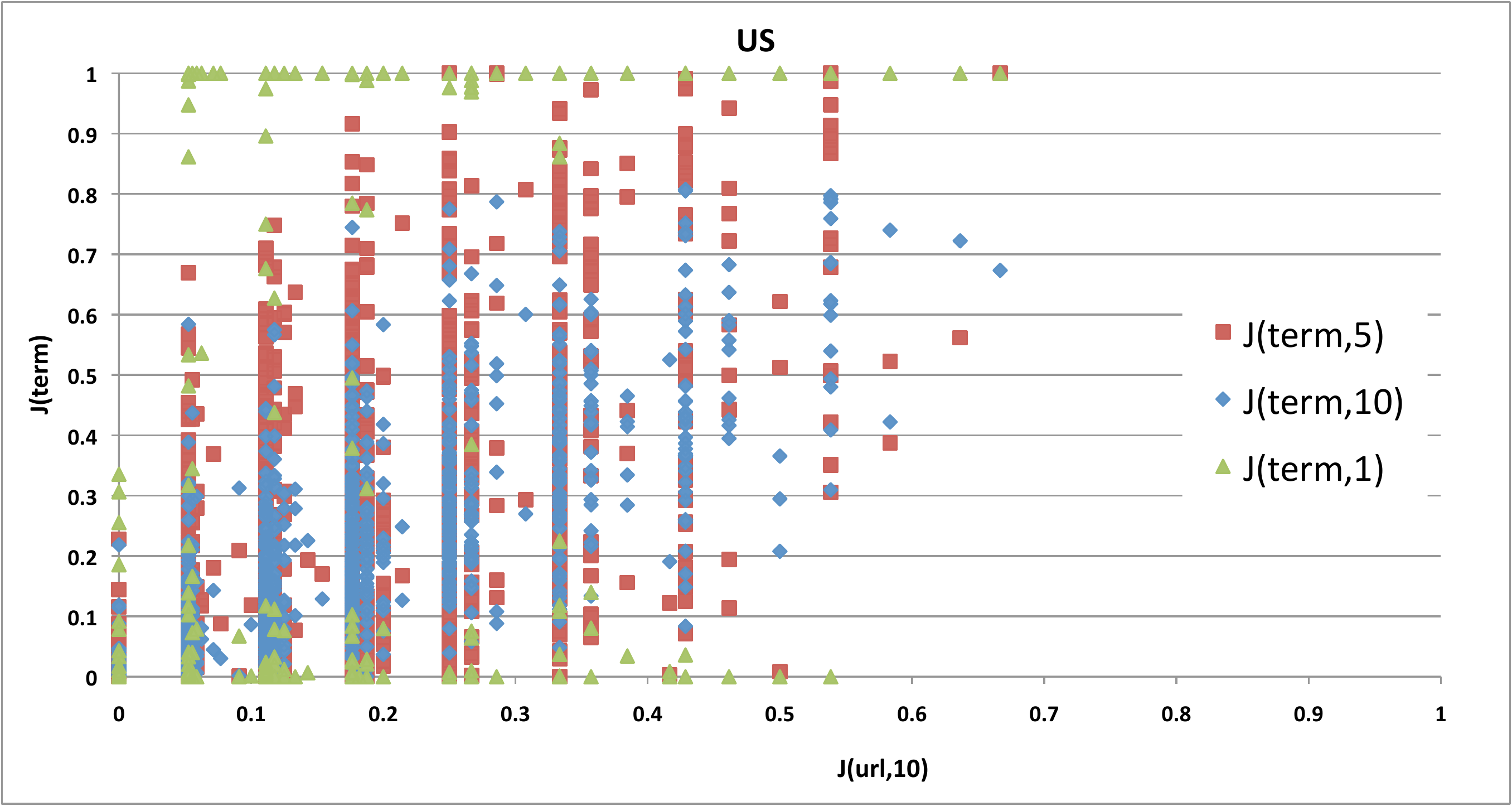}{Comparison
  of content-based and URL-based Jaccard ratios  $J_{term,n}$ and
  $J_{url,10}$ at different contents sizes. }{fig:contents-and-jaccard}

Let us refresh our memory about these contents-based measures:
URL-based Jaccard ratio is computed by first normalizing the URL name
by duplicate detection. Then the URL results are taken as list and the
intersection/union ratio is computed. The duplicate detection is
computed by using shingles or word histograms. If a threshold is
reached, then the two URLs are considered identical and only one URL
will be placed on both positions. For contents-based Jaccard ratio we
take the shingles of all documents in a results list up to a specific
rank and then we compute the intersection/union ratio of both
lists. We may summarize that the former emphasizes the discrete nature
of the list, instead the latter emphasizes the full contents of the
documents in the lists.

In Fig.~\ref{fig:contents-and-jaccard}, we show the relationship
between the content-based Jaccard ratios $J_{term,n}$ for contents from
top-$n$ search results for $n=1,5,10$ and the the URL-based Jaccard
ratio $J_{url}$ for the US and JP markets. It seems that a single
search result is too few to show similarity based on contents; that
is, taking the top results is a hit/miss measure and thus very
limited. However, if we use the first five search results we have
enough information to reach the whole similarity range
(i.e,. [0,1]). The ability to provide enough information in only the
first five results is probably because of the emphasis of search
engines to return the key results at the top. 

Notice that Fig.~\ref{fig:contents-and-jaccard} (and
Fig.~\ref{fig:footrule-and-jaccard}) show the same information about
the URL-based Jaccard presented in Fig.~\ref{fig:Jaccard-and-query}
but we did not create bins. We want to show that even thought we
wanted to collect 10 URLs per engines, there are queries for which we
have less than 10. We can see that $J_{url,n}$ is in clusters (i.e.,
close vertical lines) having the same number of common URLs but
different number of search results.

\subsubsection{Content-based measures vs. normalized footrule}

Now, we finally present the comparison between content-based measures
such as $J_{term,k}$ and $\phi_{term,k}$ versus the most common
correlation measures. The goal is to expose the different information,
presented as quantitative value by the two different types of
measures. 

In Fig.~\ref{fig:us-footrules-and-contents}, we present the
relationship between the content-based measures and the normalized
footrule for the US market ---for which we have the largest overlap in
our experiments. This is to show how small the range of the normalized
footrule is and, in contrast, how the range of the content-based
measure offers more variety and insights. It also shows that these
effects greatly magnify when the overlap is really low, which we have
shown is increasing common.

\Doublefigure{0.49}{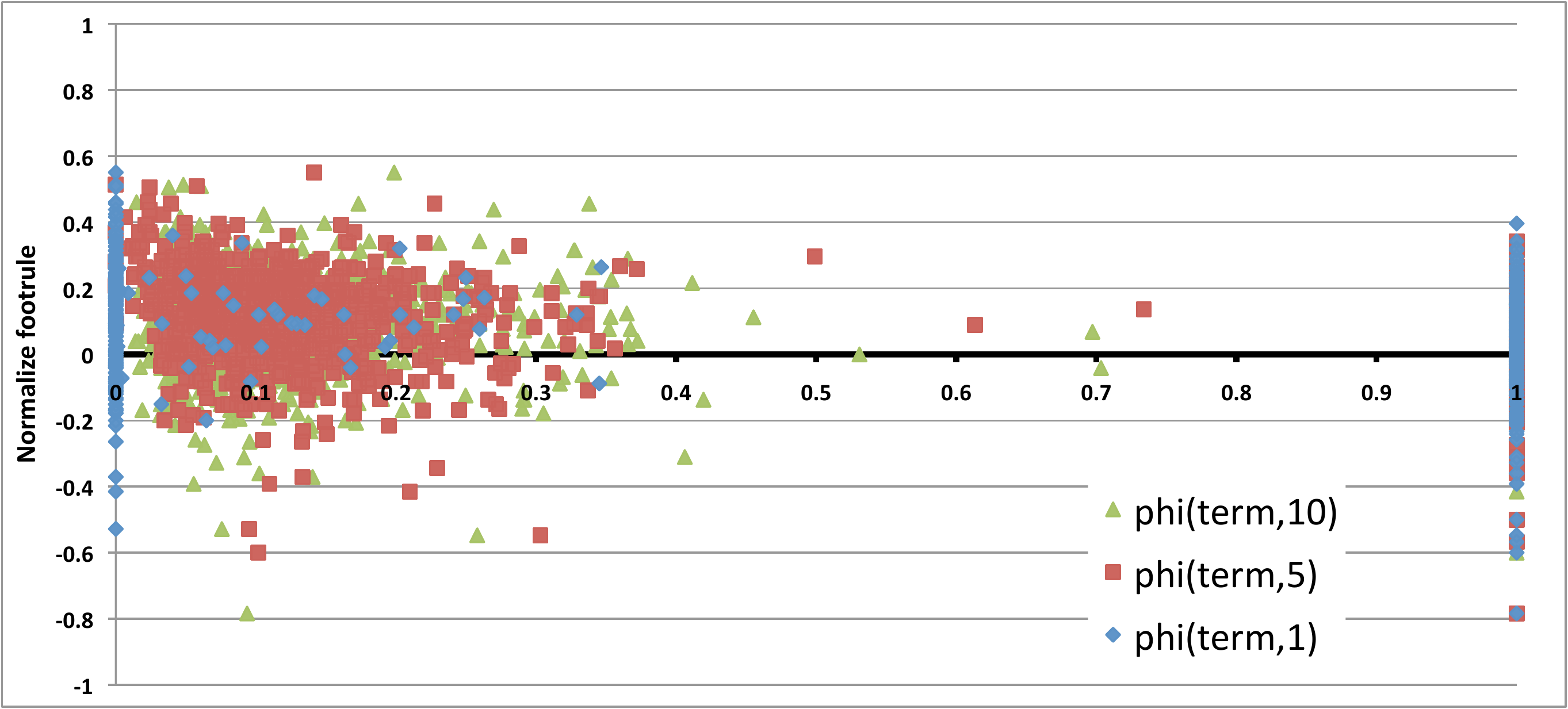}{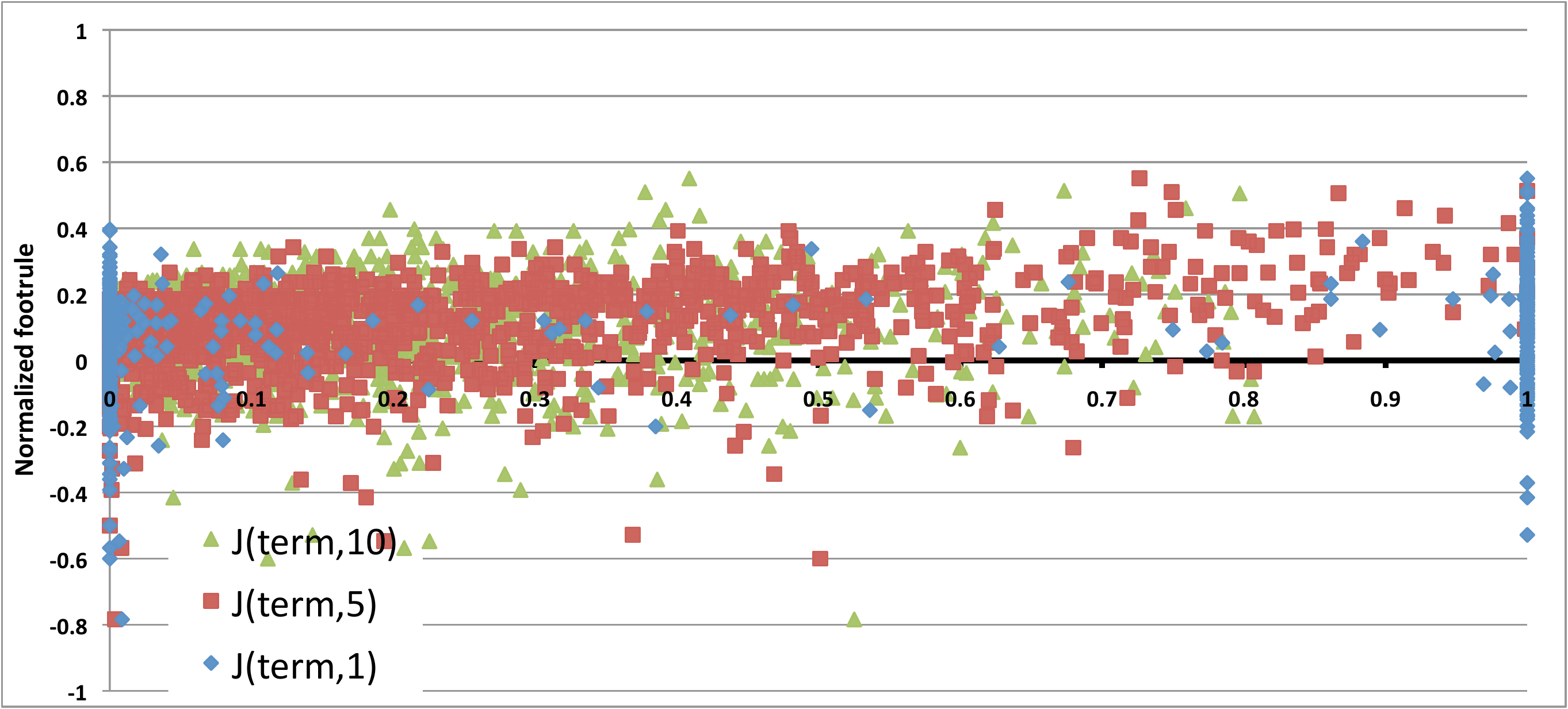}{Comparisons
  for the US market; top: the distribution measure $\phi_{term,n}$
  vs. normalized footrule $s_w(url,10)$; bottom: content-based Jaccard
  ratio $J_{term,n}$ vs. normalized footrule $s_w(url,10)$.
}{fig:us-footrules-and-contents}

We show that the normalized footrule will be indifferent to cases
where the first five documents in the lists are almost perfect
duplicates ($s_w(url,10)\sim 0$ and $J_{term,5}=1$). First, let us
recall what $J_{term,5}$ means: take an engine list and consider only
the first 5 URLs, take the contents of the documents as shingles (10
words each shingle without repetition) and create a set, and now
perform the Jaccard ratio of the sets so determined. Let us interpret
this situation: both engines give the same contents in the top of the
results, it could be the same URLs (but it is not really
important). On one side, this will provide the same experience to the
user, we see intuitively that the engines are highly correlated for
the query; on the other side, the footrule measure does not provide
any information, despite our best efforts to find common items in the
lists. In such a case, having a 10 URLs lists (20 total) is large
enough that if only 3-4 URLs are really common and high in the result
list, the footrule is dominated by the denominator and the
contribution in the numerator is mixed. As a note, the size of one
document, may dominate the $J_{term,5}$ value ---even when few
documents have common contents. This is a natural weight and, in
practice, contents based measures emphasizes the literal size of the
common documents. So we have a correlation measure for which we can
interpret the value in a more intuitive fashion and it is more
discriminative.

Let us take a look at the range of the $\phi_{term,n}$ measure and let
us recall what the measure means: take the first 5 URLs of each lists
(e.g., $n=5$), we create an histogram word--count by the contents of
the documents, then we compare the histograms by creating a cumulative
distribution function (CDF) and apply the formula Eq.\ref{eq:phi}. If
we use a lexicographical sort and a natural merge algorithm of the
words, we can always create a CDF out of the histograms.  We present
the raw distance and the function has a natural range between 0 and 2
---where 0 means equality, 2 difference, but as a function of $m$ the
number of different words the real statistically difference may be as
small as 0.2). In this figure, it seems that the function has a
limited range but for this function we have a significance value or
p-value. There are two reasons: First, we require at a minimum 30\%
overlap before to perform any comparison (from histograms to CDFs);
otherwise we state a distance of 1 and p-value of 1. Second, for this
distance function (and for all the stochastic distance function we
used in this work) we do have a statistical confidence level or
p-value, which offers further granularity for the distance measure as
described above. The footrule confidence will not adjust to the
different range, but we have reformulated the problem in such a way
that we can use a statistically sound approach with a confidence level
making this measure more discriminative and suitable for a automatic
approach (practically independent of the measure range).

 
\subsubsection{Jaccard ratio with different contents sizes}

In practice, we are introducing a correlation measure that return a
vector of values: we can compute $n$ values of the contents-based
Jaccard ratio $J_{term,n}$, here we presented three values for
$n=1,5,10$. Here, we show how to use the vector of values to find rank
correlation problems.

\Doublefigure{0.49}{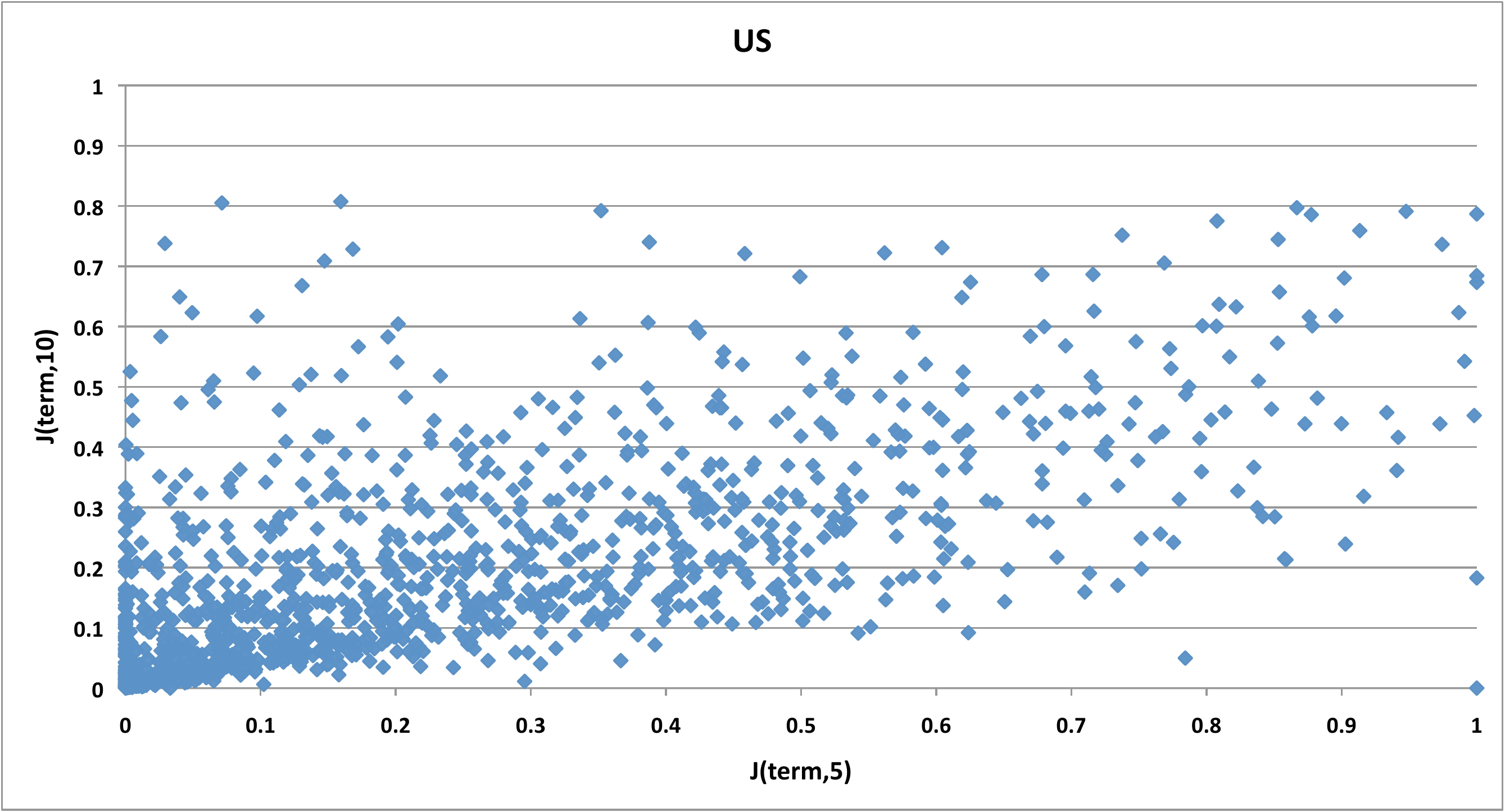}{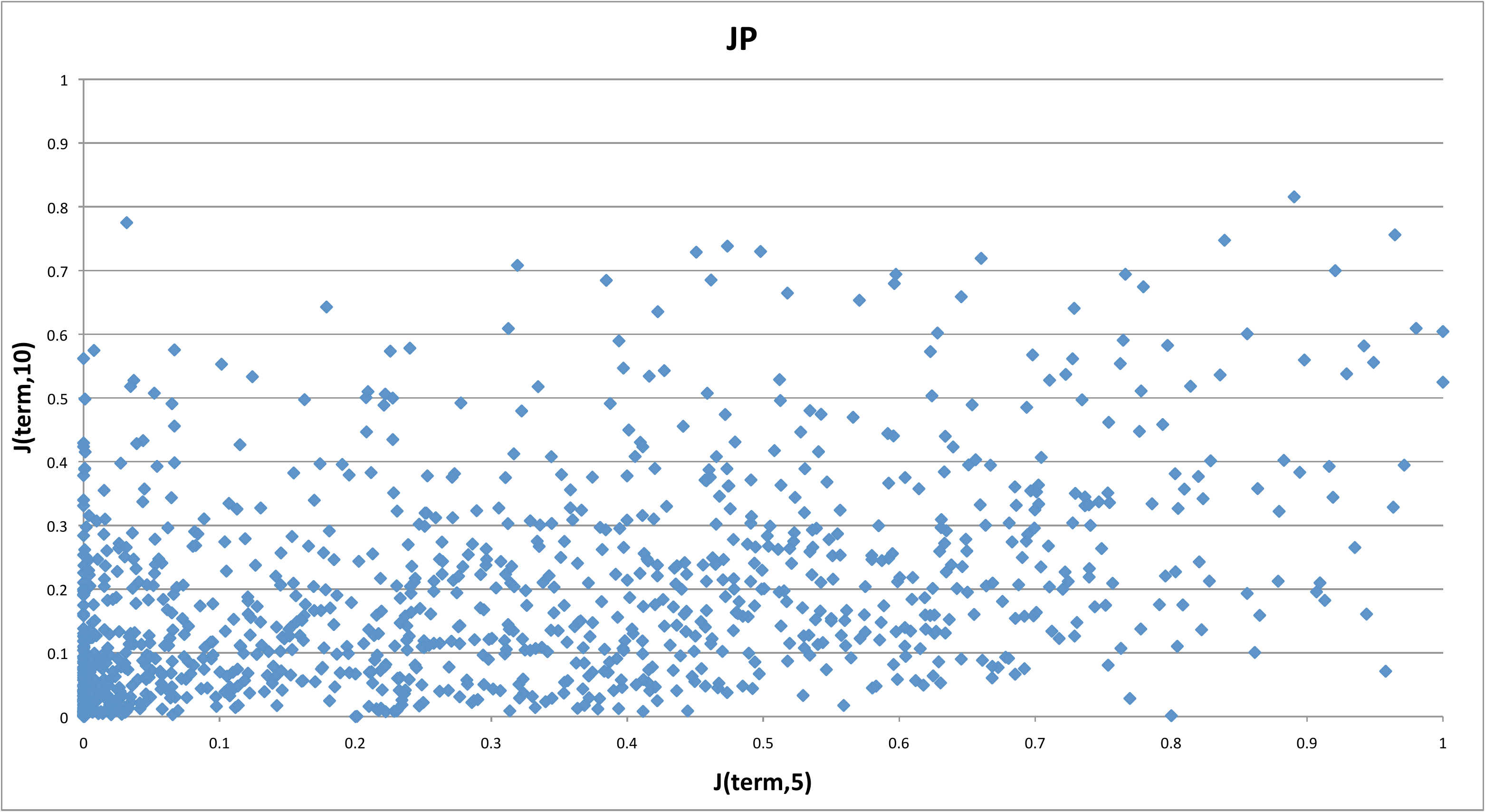}{Correlation between content-based
  Jaccard ratios with contents from top-5 and top-10 search results:
  $J_{term,5}$ vs. $J_{term,10}$; left: the US market; right: the JP
  market.}{fig:contents-and-contents}

In Fig.~\ref{fig:contents-and-contents}, we show a scattered plot for
the US and JP markets for the content-based Jaccard ratios for
different contents sizes $n=5,10$. The relatively strong correlation is
evident from these plots. Intuitively, if there is a strong
$J_{term,5}$, that is the results lists are {\em top heavy}, having
lots of common contents, this will contribute to $J_{term,10}$ as
well. 

The most interesting cases are where $J_{term,10}>J_{term,5}$, that is
the tail of the result lists are richer of common contents than the
heads. For example, with the simple rule that $J_{term,5}<J_{term,10}$
and $J_{term,10}>0.2$, we have found queries for which ranking of one
of the search engines had problems. Let us elaborate this. If
$J_{term,5}<J_{term,10}$ we can see two possible cases. First, the
tail of the results list has contents common to the head of the other,
this is the classic case of inverse correlation. Second, the tails of
both lists have the common contents, thus the heads are different,
this is a case of un-correlated results. In both cases, the queries
exploit different engine rankings. A supervised approach may take
these queries and verify whether we return the better results
(editorial test) or otherwise why our system did not return the other
engine results. Each such case provides a way to automatically
generate training data or regression tests for machine-learned ranking
systems. Think about this process of query selection as a filtering so
that only the queries requiring editorial judgment are necessary and
then can be used for training of ranking/relevance systems.

\subsubsection{Overlap by Jaccard ratio vs. results quality by DCG}

We conclude with a final evaluation of the content-based measures
($\phi_{term,10}$ and $J_{term,10}$) with the contents quality as
measured by DCG5.

\Doublefigure{0.49}{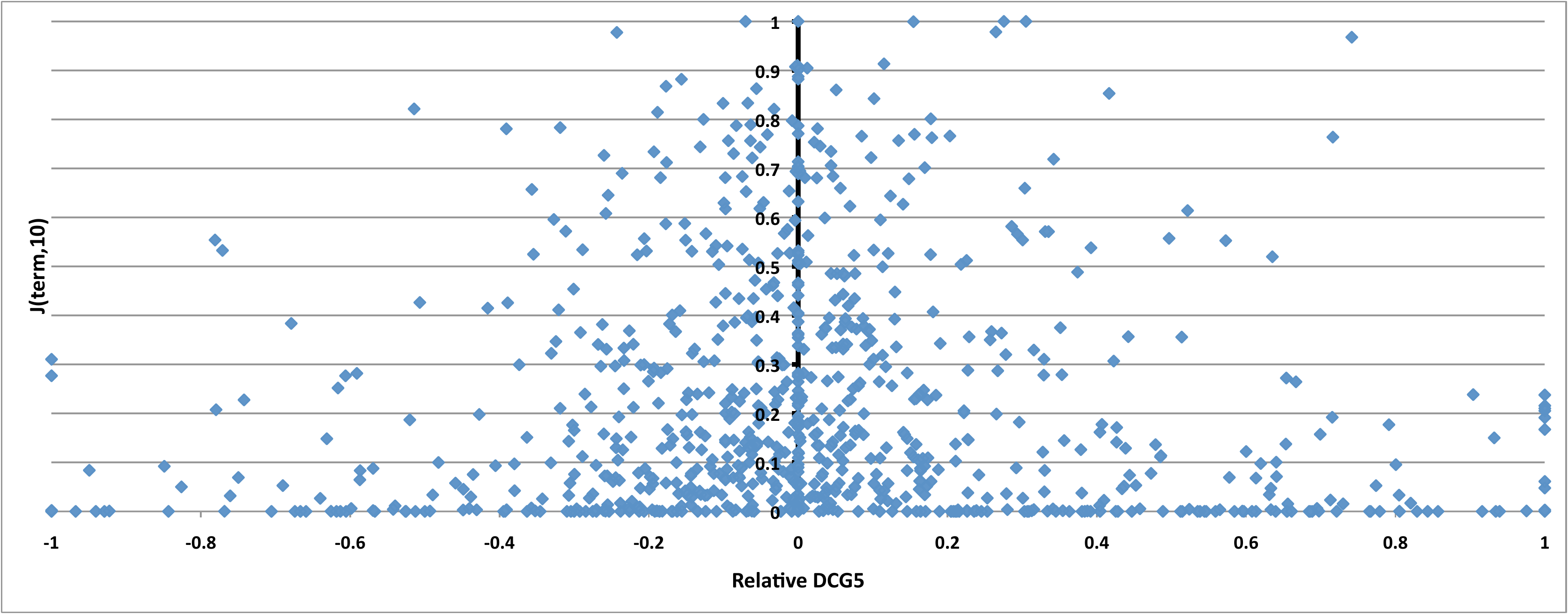}{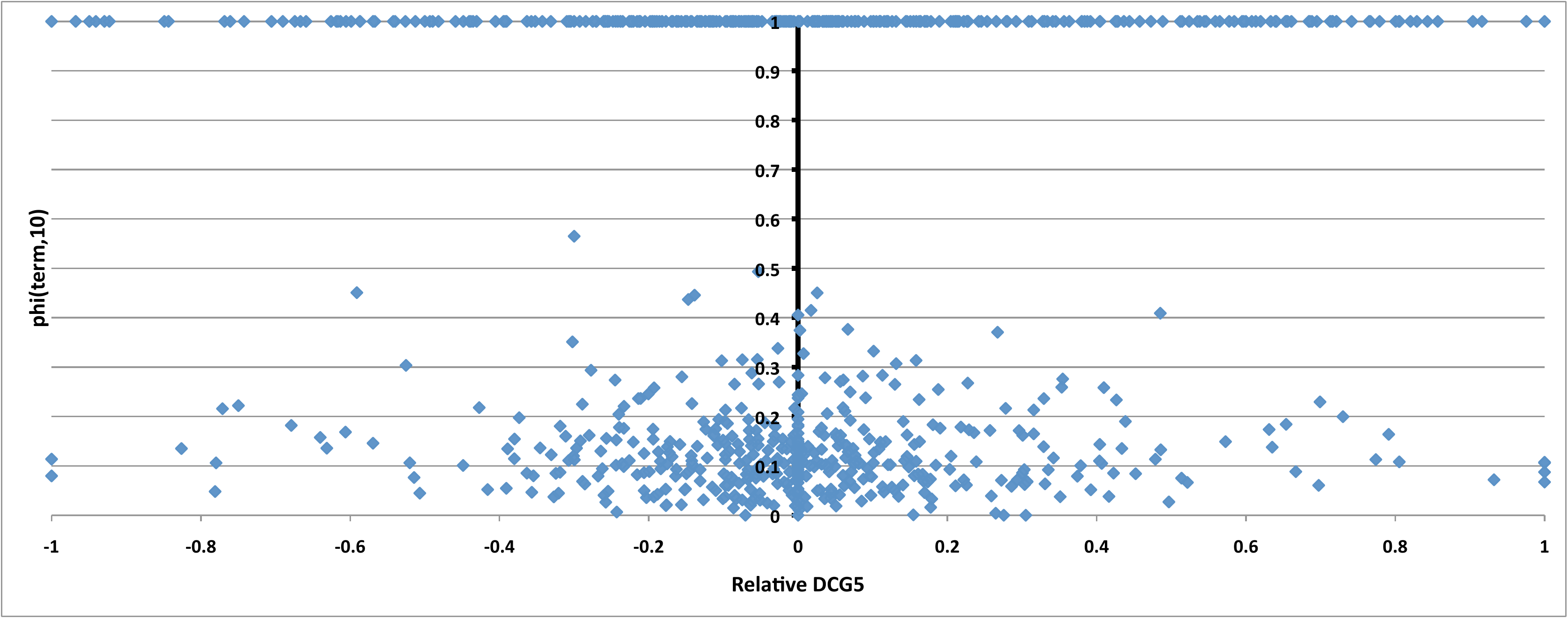}{Correlation between
content-based measures and relative DCG; left: the Jaccard ratio
$J_{term,10}$ vs. relative DCG; right: the distribution measure
$\phi_{term,10}$ vs. relative DCG.}{fig:contents-and-DCG}

We present our experimental results in Fig.~\ref{fig:contents-and-DCG}
and the conclusions are similar to what we have found previously and
presented in Fig.~\ref{fig:DCG-and-lists}: DCG5 varies greatly when
the overlap is low (URL or contents). In other words, the results
quality can cover the whole range from perfect to bad results when the
overlap between the results is low. This result also justifies that
low overlap between two major search engines does not necessarily make
one of them also better in results quality but clearly low overlap
does not mean little correlation (or inverse correlation), it means
that we can infer very little about the correlation of the results.

We would like to conclude this section and the experimental result
section noting that ---at the least--- we have presented correlation
measure that are more discriminative than the existing list-based
correlation measures for search engine results. These measures can be
certainly used as a filtering tool so that to find the queries that
really need supervised approaches or used as testing tools for the
debugging of a search engine pipeline.

\section{Conclusions}
\label{sec:conclusions}

We present how to measure search-results overlap using URL-based and
content-based measures, with contents derived from the documents at
the landing pages of the URLs in search results. We extend such
measures to carry weights and also work for permutations as well as
partial lists. In a separate and concurrent work \cite{DasdanD2011},
we prove the equivalence of the weighted generalizations of two
well-known list similarity measures.

We show that the overlap between the results of two major search
engines is fairly low (for over 80\% of the queries, no more than
three URLs). This result makes the application of URL-based measures
difficult, thereby increasing the importance and applicability of
content-based measures. We also show that low overlap does not
necessarily indicate the superiority of one search engine over another
in terms of results quality; the quality can vary greatly along the
quality range when the overlap is low.

We present many results on the sensitivity of the proposed measures to
different parameters (e.g., number of items in the lists) as well as
the relationships between the measures (list-based vs. contents-based
measures). We also briefly discuss how these measures can be used to
automatically create regression tests (i.e., filtering out query for
which two engines do well already) or training data for
machine-learned ranking systems (i.e., filtering the query that need
editorial judgment). In turn, this automatic selection of queries can
be used for the debugging of the search engine pipeline and automatic
classification could be obtained by the engineering team.

\section*{Acknowledgments}

Under the hood of this machinery, we used several components and
consulted very capable engineers: Suresh Lokia for the set of queries,
Kexiang Hu for the scraping tool, Marcin Kadluczka for the high level
fetching system for the retrieval of the documents in real time, and
Amit Sasturkar and Swapnil Hajela for the word-view pipeline and
document signature. We also thank Santanu Kolay for useful discussions
on various aspects of this work and Ravi Kumar for discussions on the
weighted form of Kendall's tau.

\section*{Acknowledgments}

Under the hood of this machinery, we used several components and
consulted very capable engineers: Suresh Lokia for the set of queries,
Kexiang Hu for the scraping tool, Marcin Kadluczka for the high level
fetching system for the retrieval of the documents in real time, and
Amit Sasturkar and Swapnil Hajela for the word-view pipeline and
document signature. We also thank Santanu Kolay for useful discussions
on various aspects of this work and Ravi Kumar for discussions on the
weighted form of Kendall's tau.

\bibliographystyle{plain}
\bibliography{biblio}
\appendix{Reviewers' Comments}
The community has spoken about and against this work. Here we share
the anonymous considerations without our reply. Enjoy the drama. 

\appendix{Reviewers' Comments Journal 1}

Dear Paolo,

Thanks for asking. Unfortunately, after having tried quite a few
potential ---- reviewers, we are not able to get even one referee
report. Most of them declined to review, and some of them suggest this
paper is not well within the scope of ----. The guardian editors of
this work have evaluated the situation, they are convinced this paper
is most likely not interesting to ---- readers, by looking at
especially the people and journals/conferences mentioned in their
related work. they consider this work is more web search than web
engineering.

Therefore, it should be the best interests of the authors to find some
other better suitable journal to this work. We return this paper back
to you as the author, and wish you good luck somewhere else.

Regards, Wei for ---- Editorial

\appendix{Reviewers' Comments Journal 2}

\mypar{Second Round}

Dear Dr. Paolo D'Alberto:

We have received the reports from our advisors on your manuscript, "On
the Divergence of Search Engines' Results (Unsupervised Comparison of
Search Engine Rankings)".

With regret, I must inform you that, based on the advice received, the
Editor-in-Chief has decided that your manuscript cannot be accepted
for publication in World Wide Web Journal.

Attached, please find the reviewer comments for your perusal.

I would like to thank you very much for forwarding your manuscript to
us for consideration and wish you every success in finding an
alternative place of publication.

Comments for the Author:

Reviewer 2: The paper addressed most of reviewers' comment reasonably
well. Presentation has been improved greatly: scoping and motivation
of the problem has been substantially improved, and it's now in a good
shape. The impact of the paper remains at the same level: not as
strong as groundbreaking, but a useful proof plus empirical studies on
the weakness of list-based comparison methods, and also
suggestion/validation of a content-based comparison method.

The reviewer recommends the paper for the publication in ----, after
the minor revisions discussed below:

1) section 3.2.1
the same as they are in the original lists \verb2\2sigma and \verb2\2pi, respectively. -->
the same as they are in the original lists \verb2\2pi and \verb2\2sigma, respectively.

2) same section 3.2.1 example
(a,b,c,e,f) -->
(a,b,d,e,f)

3) question: why use a,b,d,e,f? why no "c"? It's not even an issue,
but just curious...

4) section 3.2.2 example (nice example, BTW) it's not clear how $s_w$
(normalized version) denominator is computed in the example. $S_w$ has
been shown to the detail, and it'll be nice to show the same procedure
for the denominator (so that the reader doesn't have to wonder.) Also
it seems that $s_w$ uses $2S_w$ rather than $S_w$ at the top, so
shouldn't it be $20w$, instead of $10w$? Also, $s_w$'s $w$ can be
cancelled form the top and bottom, so two $w$'s should cancel each
other?

5) Figure 1. For the same countries, "JP" and "US", it'll be nice to
use the same color.

Reviewer 3: My primary complaints on the earlier draft were that

(1) The set (or list) similarity section is marginally related to the experimental part of the paper.
(2) The contribution and the conclusion of experimental section were not clear.
(3) The paper is difficult to follow at various places and needs significant revision.

In their reply, the authors tried to make the case for the relevance
of their similarity part, but I am still not convinced. There are no
new insights or results that the authors added to the new draft of the
experiment section. The writing of the paper has improved but it still
needs to be polished more. Based on these, I recommend rejecting the
paper.

Here are more detailed comments.

(1) The authors argue that the similarity-metric equivalence result is
significant because it gives credibility to their results in the
experiment section. I do not agree with this argument. What is new in
the paper (in terms of the similarity metric equivalence) is their
extension of the equivalence theorem to weighted metrics. The
equivalence of unweighted metrics are already known in the
literature. Unfortunately, in their experimental section, the authors
eventually decide that they will use only unweighted metrics. Then
what was really the point of Section 3? Why do you need to prove the
equivalence of weighted metrics when you do not use them?

(2) In the original review, I complained about the significance of
results reported in the experimental section and the difficulty of
reading parts of the section. The writing quality of the experimental
section has improved in the new draft, but no new results or insights
have been added. I am still not clear about what is the takeaway
message of the results reported in the experimental section.

(3) At many places, the paper still needs quite a bit of proof reading
and/or polishing. I will point out problems in Sec 3.2 as an example:

(a) Line 42 of Sec 3.2.1: pi(i) = n + sigma(i) - 1: what is n here?
Since we are dealing with partial list, the meaning of n is different
from earlier definition of n. I also believe that the equation should
not have -1 at the end. Assuming n is the length of pi, when we append
an element at the end of pi, its rank starts with n+1, not n.

(b) Line 58-61 of Sec 3.2.1: I do not understand this statement.

(c) Equation (5). the denominator has (n-i+1). Again, I am not clear
what n means here.

(d) equation on s\verb2_2w after Equation (6). I am not sure why it simplfies
to 1 - w 1/3. I also do not see why the denominator grows as $n^2$.

(e) Equation (7). sigma = iota(?). Iota has not been defined.

(f) Line 31 on the right column of 5. What is F metric?

The paper has errors like these in other parts as well, which make it
difficult to follow.

Reviewer 4: Second review of "On the divergence of search engines'
results" The results of Section 3 seem to be unrelated to those in
later sections.  There is some improvement in presentation, but
further improvement is needed.  Some examples: p.4 first example,
sigma prime "c" replaced by "d"?  How does the normalized Kw become
negative?  p.6 Example phi(Fsigma, Fpi) =2. The two distributions are
not that different. Why the maximum difference?  p.8 an example in
Section 5.3.1 would help.

Sections 6.3.2- 6.3.4 need to be presented better. The figures are
hard to read with three figures superimposed together. Better
explanations should be provided.

\mypar{First Round}

Dear Dr. Paolo D'Alberto:

We have received the reports from our advisors on your manuscript, "On
the Divergence of Search Engines' Results (Unsupervised Comparison of
Search Engine Rankings)", which you submitted to World Wide Web
Journal.

Based on the advice received, the Editor feels that your manuscript
could be reconsidered for publication should you be prepared to
incorporate major revisions.  When preparing your revised manuscript,
you are asked to carefully consider the reviewer comments which are
attached, and submit a list of responses to the comments.  Your list
of responses should be uploaded as a file in addition to your revised
manuscript.

COMMENTS FOR THE AUTHOR:

Reviewer 1: The paper proposes a method for comparing the results
from different search engines. The paper is well motivated and has a
potential of practical use. However, the first contribution claimed,
proof of equivalence between a weighted generalizations of Spearman's
footrule and Kendall's tau, is weak since it is a simple extension of
the existing work, the proof for the unweighted permutations
[9]. Moreover, I'm not sure whether the proof should be included in
this paper. It consumes much space but it is not essential part of the
paper. It would be enough to choose one of two measures. The other
contributions claimed are applications of the existing work. It is
difficult to find significant technical contributions.

Reviewer 2: The authors claim three main contributions - i) proof for
equivalence of extended version of Spearman's footrule and Kendall's
tau, ii) observation of divergent results from multiple search
engines, and iii) content-based similarity measurement of search
engine results.

First contribution appears to be a solid and useful contribution that
can be used for general similarity measurement methods. However, to
the reviewer, it seems that the rest of the paper is not very strongly
motivated - why should readers care about the divergence of search
engines? The current status of art provides a reasonable quality, and
the fact that different search engines produce different results is
hardly surprising considering the scale of web and the difficulty of
search task. The paper may be interesting for some engineers at
Google, Yahoo, or Microsoft, but to the general audience, it's not
clear what they gain from the paper. Authors recommend that users
should use meta-search or multiple search engines because of the
divergence, but it seems that users are fine with what they get from a
single search engine, and the suggestion doesn't seem to make sense.

One possible direction for improvement is to discuss more about the
detailed anlysis of search engine biases, such as which search engine
is good at what, and not so good at what, rather than simply reporting
that they are different. This may give general audiences a better
insight toward the current status of multiple search engine
technologies.

Reviewer 3: Summary:

In this paper the authors present URL-based and content-based measures
of search engine result overlap.  For URLs, they show that (1) even
with normalized URLs, overlap (e.g. Jaccard ratio) is generally small,
and (2) URL overlap is not indicative of quality.  Given this, they
suggest content-based approaches for measuring overlap. They show
content-based approaches (e.g. shingle overlap from the top n pages)
provides a wider range of values, and again no correlation between
overlap and quality.  These apply to both ordered (set) and unordered
(list) measures.  These measures can be used, for example, to see
where one search engine performs better than another.

Comments:

* The bulk of the contributions seems to be Section 3 and small
observations about the figures throughout Section 6.  Could use more
insight into what each of the graphs really means.  Isn't the "second"
contribution really just motivation for the use of content-based
comparisons?

* Very long discussion and proofs of footrule and Kendall's, their
equivalence, etc in Section 3... but then the figures seem to suggest
footrule is not particularly useful for comparing search results
anyways (due to high divergence)?  So then why is their equivalence or
the extension to weighted lists important here?

* A lot of the details of the paper seem to be in areas mostly
unrelated to what I perceived as the main point (most of Section 3,
5.1, 5.3).

* Query classification (e.g. navigational) could make a huge
difference on the results.  One would reasonably expect navigational
queries to have much higher URL correlation than, say, informational
queries, particularly in the top few results.  The high J(term,1) and
phi(term,1) results in Fig 4 and 6 could be due to this, and it could
raise the overall content-based similarity scores.

* The paper could benefit from reorganization.  Motivations aren't as clear upfront (e.g. at the time, I didn't really know why I was reading through the messy details of Section 3). The "Normalization" steps in Section 5.3.1 is almost unreadable.  I'm not clear what this is saying or what each of the symbols really means.

* Are the x-y labels for Fig 7 (JP) correct?

Reviewer 4: The paper has three results. (1) a proof that Spearman's
footrule and Kendall' tau are equivalent. (2) Most queries have very
little overlap for the top two search engines, Google and Yahoo, in
their top 10 results. (3) Introduce measures to compare performance of
search engines based on contents. The results are somewhat
interesting, but the presentation needs improvement. There is not even
a single example in the entire paper. The authors should utilize
examples to illustrate their ideas.  

Specifice comments: 

p.5 Give the intuitive idea of Kw in equation (6).  line after
equation (8) Should it be the numerator instead of the denominator?
SMetric space: Should F be Section 3.2.4 Equivalence usually implies
"stronger than within small constant multiples of one another"?

p.7 Section 4.4 second para What is exactly the similarity score
defined in [22]?  

p.8 left line 32 Should the Jaccard ratio be above 0.5?  Section 5.3.1
I am lost. What is sigma zero? What is the intuition for
normalization. Please explain Step 1 and step 2 clearly.  

p.9 right l.48 Why are "the search engines seem correlated"?  

p.10 first para Fig. 3 The values in the range [0.4, 1] seem to be
larger than those in [-0.4, -1]. Does'nt that imply one search engine
has better performance than the other?  

p.10 I don't understand what Fig. 4 shows. Please explain clearly.
What exactly are the purposes for detecting near-duplicates using
shingles? Is it used to detect near duplicates among documents in the
search result of one search engine and those in the search result of
the other? Or, the near duplicates are detected among documents within
single search engine?  

p.11 left lines 31-33 I have difficulty understanding this sentence.
l.57 What is meant by "we could create queries for which ranking pf
one of the search engines had problems" and how?  Section 6.3.5 For
the statement "DCG5 varies greatly when the overlap is low", should'nt
DCG5 be the Y-axis and the overlap be the X-axis?  Conclusion It is
not clear "how these measures can be used to atomatically create
regression tests or training data for machine-learned ranking
systems"?

\appendix{Reviewers' Comments Conference 1}
to add

\end{document}